\documentclass[amssymb,aps]{revtex4}
\emergencystretch=\maxdimen
\usepackage{graphicx}
\usepackage{dcolumn}
\usepackage{amsmath}
\usepackage[latin1]{inputenc}
\usepackage{amssymb}
\usepackage[colorlinks=true, citecolor=blue, urlcolor = blue, linkcolor= red, bookmarks=true]{hyperref}
\usepackage{float}
\usepackage{amsfonts}
\usepackage{dcolumn}
\usepackage{hyperref}
\usepackage{subfigure}
\usepackage{pgfplots}
\usepackage{booktabs}
\usepackage{mathrsfs}
\usepackage{multirow}
\pgfplotsset{compat=1.16}

\usepackage{adjustbox}

\makeatletter
\def\btt#1{\texttt{\@backslashchar#1}}
\DeclareRobustCommand\bblash{\btt{\@backslashchar}} \makeatother

\begin{document}

\title[]{Strong field gravitational lensing by hairy Kerr black holes}
\author{Shafqat Ul Islam$^{a}$ } \email{Shafphy@gmail.com}

\author{Sushant~G.~Ghosh$^{a,\;b}$} \email{sghosh2@jmi.ac.in, sgghosh@gmail.com}

\affiliation{$^{a}$ Centre for Theoretical Physics, 
	Jamia Millia Islamia, New Delhi 110025, India}
\affiliation{$^{b}$ Astrophysics and Cosmology Research Unit, 
	School of Mathematics, Statistics and Computer Science, 
	University of KwaZulu-Natal, Private Bag 54001, Durban 4000, South Africa}

\begin{abstract}
Recent times witnessed a surge of interest in strong gravitational lensing by black holes due to the Event Horizon Telescope (EHT) results, which suggest comparing the black hole lensing in general relativity and modified gravity theories. This may help us to assess the phenomenological differences between these models.  A Kerr black hole is also a solution to some alternative theories of gravity, while recently obtained modified Kerr black holes (hairy Kerr black holes), which evade the no-hair theorem,  are due to additional sources from surrounding fluid,  like dark matter, having conserved energy momentum tensor (EMT).  These hairy Kerr black holes may also be solutions to an alternative theory of gravity.   We generalize previous work on gravitational lensing by a Kerr black hole in the strong deflection limits to the hairy Kerr black holes, with a deviation parameter $\alpha$ and a primary hair $\ell_0$.  Interestingly, the deflection coefficient $\bar{a}$ increases and decreases with increasing  $\ell_0$ and  $\alpha$ respectively. $\bar{b}$ shows opposite behaviour with $\ell_0$ and  $\alpha$.  We also find that the deflection angle $\alpha_D$, angular position $\theta_{\infty}$ and $u_{m}$ decrease, but angular separation $s$ increases with $\alpha$.  We compare our results with those for Kerr black holes, and also  apply  the formalism to discuss the astrophysical consequences in the context of the supermassive black holes Sgr A* and M87*. We observe that the deviations of the angular positions from that of the Kerr black hole are not more than $2.6~\mu$as for Sgr A* and $1.96~\mu$as for M87*, which are unlikely to be resolved by the current EHT observations. 
\end{abstract}

\pacs{04.50.Kd, 04.20.Jb, 04.40.Nr, 04.70.Bw}

\maketitle
\section{Introduction}
In Einstein's general relativity (GR), there is only one uncharged rotating black hole solution given by the Kerr metric \cite{Kerr:1963ud}. The no-hair theorem states that black holes are uniquely characterized by their mass $M$, and spin, $J$, and are described by the Kerr metric \cite{Israel:1967,Carter:1971,Robin:1975}. Consequently, all astrophysical black holes are expected to be Kerr black holes. A rotating non-Kerr metric black hole has an additional deviation parameter from modifying gravity or matters, apart from mass and rotation parameters, and encompasses the Kerr black hole as a particular case \cite{Kumar:2020yem}. While there is some indirect evidence suggesting that the Kerr metric  has an event horizon, they are black holes; a proof that the Kerr geometry describes the spacetime around these objects is still lacking, and it may be difficult to rule out non-Kerr  black holes \cite{Ryan:1995,Will:2006}. Rotating regular black holes \cite{Bambi:2013ufa,Azreg-Ainou1:2014pra,Ghosh:2014pba}, prototype non-Kerr black holes have also been obtained and studied extensively, which in the large $r$ limits retrieve the Kerr black hole solution \cite{Kerr:1963ud}. An important question arises: are such black holes candidates for testing the no-hair theorem or the Kerr hypothesis? The Kerr hypothesis, a strong field prediction of GR, may or may not hold for the non-Kerr black holes \cite{Kumar:2020yem}. Here, we use gravitational lensing as a tool to investigate the constraints when rotating non-Kerr black holes can be considered as astrophysical black hole candidates.  The gravitational decoupling approach (GD) \cite{Ovalle:2017fgl,Ovalle:2019qyi} is precisely designed for describing deformations of known spherically symmetric solutions of GR induced by additional sources. This method is useful for generating new and more complex solutions from known (seed) solutions of the Einstein field equations and  modified gravitational theories.   Recently,  it is shown that the GD approach could be used to obtain axially symmetric systems \cite{Contreras:2021yxe}. Indeed the black hole solution contains a source satisfying the strong energy condition (SEC) and provides a modification of the Kerr metric termed as Kerr black holes with primary hair by  Contreras \textit{et. al.} \cite{Contreras:2021yxe}. In GR, there are black holes with hair due to global charge (see \cite{Herdeiro:2015waa} for a review). In general, hairy black holes are referred to as stationary black hole solution with new global charges or new non trivial fields which are not associated with Gauss law \cite{Herdeiro:2015waa}, e.g., black holes with scalar hair \cite{Herdeiro:2014goa,yaun:2021} or proca hair \cite{Herdeiro:2016tmi}.  

Deflection of a light ray in a gravitational field is referred to as gravitational lensing, and the object causing a deflection is called a gravitational lens. Gravitational lensing by black holes is one of the most powerful astrophysical tools for investigating the strong field features of gravity.  It could provide a profound test of modified theories of gravity in the strong field regimes \cite{Bekenstein:1994,Eiroa:2006,Sarkar:2006,Chen:2009,Kumar:2020sag,Islam:2020xmy} and also the cosmic censorship hypothesis \cite{Virbhadra:2002,Virbhadra:2008}. 
The strong field limit gravitational lensing studies due to black holes have received considerable attention in recent years, indicating that one can extract the black hole information from it.   Gravitational lensing by black holes began to be observationally crucial in the 1990s, which motivated several quantitative studies of the Kerr metric caustics \cite{Rauch :1994qd,Vazquez:2003zm, Bozza:2008mi,Bozza:2009yw}. Vazquez and Esteban \cite{Vazquez:2003zm} explored the phenomenology of strong field gravitational lensing by using a Kerr black hole. They have developed a general procedure to calculate the positions and magnification of all images for an observer and source far away from the black hole and at arbitrary inclinations \cite{Vazquez:2003zm}.  Since then, gravitational deflection of light by rotating black holes has received significant attention due to the tremendous advancement of current observational facilities \cite{Ghosh:2020spb,Wei:2011nj,Beckwith:2004ae,Hsiao:2019ohy,Kapec:2019hro,Gralla:2019drh,James:2015yla}. The hairy Kerr black holes might have the interesting feature in contrast to the Kerr black holes \cite{Cunha:2019ikd,Cunha:2019dwb,Cunha:2016bpi,Cunha:2015yba}. This may help us to understand the hairy Kerr black holes in a better way.  

Recent time witnessed a flurry of interest in strong gravitational lensing by black holes due to the Event Horizon Telescope (EHT) observations \cite{Akiyama:2019cqa}.  This paper aims to investigate the strong gravitational lensing  of recently derived hairy Kerr black holes \cite{Contreras:2021yxe} and assess the phenomenological differences with the Kerr black holes.  The purpose of this paper is to examine the role of the deformation parameter $\alpha$  and the primary hair $\ell_0 $ on gravitational lensing observables and time delay between the relativistic images.  Further, considering the supermassive black holes Sgr A* and M$87^*$ as the lens, we obtain the positions, separation, magnification, and time delay of relativistic images.  Our results show that there is a significant effect from the primary hair on the strong gravitational lensing.

The paper is organized as follows. In  Sec. \ref{gdapproach}, we briefly review the recently obtained hairy Kerr black holes. A formalism for gravitational deflection of light in the strong field limit is the subject in Sec. \ref{sgl}.  In Sec. \ref{observables},  we discuss the strong lensing observables by hairy Kerr black holes including the positions, magnifications, and time delays of the images. Interestingly, by  taking the supermassive black holes   Sgr A* and M87* as the lens, we numerically estimate the observables  in Sec. \ref{cal}. We conclude with our significant results in Sec. \ref{conclusion}. Throughout this paper, unless otherwise stated, we adopt natural units ($G\; =\; c\;=\; 1$).

\section{The GD approach for Hairy Kerr black holes}\label{gdapproach}
Recently, Ovalle \textit{et.al.}~\cite{Ovalle:2020kpd} (see also \cite{Ovalle:2017fgl,Ovalle:2019qyi}), proposed  a simple approach 
to generate spherically symmetric hairy black holes by requiring a well-defined event horizon and the SEC or dominant energy condition (DEC) for the hair outside the horizon, which they extended to the rotating case \cite{Contreras:2021yxe}. Throughout the paper we shall call the procedure to generate the deformed solutions  the  GD approach.  We briefly review the  GD approach to generate hairy rotating black holes.  The straightforward method is designed to create  deformed solutions to the known GR solution, because of the additional sources. Thus, by using the GD approach one has a systematic and straightforward strategy to extensions of axially symmetric black holes as well \cite{Contreras:2021yxe}. Therefore, one can obtain without much effort the Kerr black hole's  nontrivial extensions that can support primary hair \cite{Contreras:2021yxe}. Let us start with the Einstein field equations,
\begin{equation} \label{corr2}
\tilde G_{\mu\nu} = k \tilde{T}_{\mu\nu} =k(T^{\rm}_{\mu\nu}
+ S_{\mu\nu})
\end{equation}
where $T^{\rm}_{\mu\nu}$ corresponds to the energy momentum tensor (EMT)  of the known solution in GR and $S^{\rm}_{\mu\nu}$ is the EMT of the additional source \cite{Ovalle:2017fgl,Ovalle:2019qyi}. Consider a generic extension of the Kerr black hole which  in the Boyer-Lindquist coordinates is given by \cite{Bambi:2013ufa,Toshmatov:2017zpr,Kumar:2020}:
\begin{eqnarray}\label{kerrex}
ds^2 & = & - \left[ 1- \frac{2r\,\tilde{m}(r)}{\Sigma} \right] dt^2 +
\frac{\Sigma}{\Delta}\,dr^2 + \Sigma\, d \theta^2- \frac{4ar\,\tilde{m}(r)
}{\Sigma  } \sin^2 \theta \, dt \; d\phi \nonumber
\\ & + & \left[r^2+ a^2 +
 \frac{2a^2 r\, \tilde{m}(r)   }{\Sigma} \sin^2 \theta
\right] \sin^2 \theta \, d\phi^2,
\end{eqnarray}
where $\Sigma = r^2 + a^2 \cos^2\theta$,~~ $\Delta=r^2 + a^2 - 2r\;\tilde{m}(r)$,~~ $a \,=\,L /M $, and  $L$ is the angular momentum. Equation~\eqref{kerrex}  can be used to describe rotating compact objects like black holes, which encompass well-known Kerr black holes as special cases when  $\tilde{m}(r)=M$. For $a=0$, we obtain the following spherically symmetric  static metric: 
\begin{eqnarray}\label{metric}
ds^2 & = & - \left[ 1- \frac{2\tilde{m}(r)}{r} \right]   dt^2 +  \left[ 1- \frac{2\tilde{m}(r)}{r} \right]^{-1}dr^2 + r^2 d\Omega^2. 
\end{eqnarray}
In the GD approach, by deforming the spherically symmetric static black hole solution of GR one can generate rotating black hole spacetimes, e.g., one can obtain nontrivial extensions of the Kerr black holes or the hairy Kerr black holes. Let us suppose that the $\tilde{m}(r)= m(r)$ corresponds to the EMT $T^{\rm}_{\mu\nu}$ alone and adding the additional sources $S_{\mu\nu}$ leads to
\begin{equation}\label{m}
\tilde{m}(r)=
m(r)+\alpha m_s(r), 
\end{equation}
where $\alpha$ is deformation parameter. Thus, the mass functions $m$ and $m_s$ are, respectively, generated by the EMT $T_{\mu\nu}$ and $S_{\mu\nu}$. The $S_{\mu\nu}$ represents  additional sources surrounding the black hole which could be  dark matter or dark energy. The Einstein tensor has only  linear derivatives of the mass function $\tilde{m}(r)$, and hence  we also have a linear decomposition of the Einstein tensor,
\begin{equation}\label{decoA}
\tilde G_{\gamma}^{\sigma}(\tilde m, a)=
G_{\gamma}^{\sigma}(m, a) + 
\alpha G_{\gamma}^{\sigma}(m_s, a).
\end{equation}
provided the rotational parameter $a$ does not change. The Equation~(\ref{decoA}) is the requirement to generate rotating black hole solutions.
 
As an immediate consequence of the GD approach, one can generate the well-known Kerr-Newman solution of the Einstein-Maxwell system. For this we have to choose ${T}_{\mu\nu}=0$ and 
\begin{equation} \label{max}
S_{\mu\nu} = \frac{1}{4\pi}
\left(F_{\mu\alpha} F^{\alpha}_{\nu}
+\frac{1}{4} g_{\mu\nu} F_{\alpha\beta} F^{\alpha\beta} \right)
\end{equation} 
Solving the Einstein equations in the vacuum $T_{\mu\nu}=0$, we find the Schwarzschild solution with mass  $\tilde{m}(r) = M$ and for  the source $S_{\mu\nu}$ one gets the  Reissner-Nordstr\"{o}m solution, whose mass function is identified as 
\begin{equation}
\label{mM2}
m_s(r) = C-\frac{Q^2}{2 r},
\end{equation}
where $C$ and $Q$ are integration constants and  $Q$ is identified as the electric charge. Using $m(r)=\cal M$ and  Equation~(\ref{m}) gives the total mass function 
\begin{eqnarray} \label{mM3}
\tilde{m}(r) = 
{\cal M}-\frac{Q^2}{2r},
\end{eqnarray}
with ${\cal M}=C+M$. Finally, for the rotating solution we substitute Equation~(\ref{mM3}) in  metric~\eqref{kerrex}, which gives
\begin{equation}
 \Delta_{KN} = r^2 -2 r {\cal M} +{a}^{2} +Q^2.
\end{equation}
The metric~(\ref{kerrex}) with mass function~(\ref{mM3}) and the above $ \Delta$, is the well known Kerr-Newman  black hole solution.

Next, we consider the Schwarzschild black hole  surrounded by  spherically symmetric matter with a conserved EMT $S_{\mu\nu}$ satisfying the SEC. It leads to the hairy  Schwarzschild black hole \cite{Contreras:2021yxe},
\begin{eqnarray}\label{strongBH}
-\left[ 1- \frac{2\mathcal{M}}{r} +\alpha e^{-r/(\mathcal{M}-\ell_0/2)}\right]dt^2 + \left[ 1- \frac{2\mathcal{M}}{r} +\alpha e^{-r/(\mathcal{M}-\ell_0/2)}\right]^{-1} + r^2d\Omega^2, 
\end{eqnarray}
where $\ell_0=\alpha\,\ell$ measures the increase of entropy
caused by the hair and must satisfy $ \ell_0 \leq 2{\cal M}
= \ell_{\rm K}$ to ensure asymptotic flatness. 
Equation~(\ref{strongBH}) can be  used as a seed metric to generate rotating black holes. As above, identifying mass  from the metric~(\ref{strongBH}) \cite{Contreras:2021yxe} we have 
\begin{equation} \label{massBH}
\tilde{m}(r)
={\cal M}-\alpha\frac{r}{2} e^{-r/(\mathcal{M}-\ell_0/2)},
\end{equation}
which implies
\begin{equation} \label{Delta2}
\Delta =
r^2+a^2-2r{\cal M} + \alpha r^2 e^{-r/(\mathcal{M}-\ell_0/2)}.
\end{equation}
Then metric~(\ref{kerrex}) with mass ~(\ref{massBH}) and the above $\Delta$ represents hairy Kerr black holes \cite{Contreras:2021yxe}.  Finally, the spherically symmetric metric with ~\eqref{strongBH}
satisfies the SEC ~\cite{Ovalle:2020kpd}
which is  also obeyed by the rotating metrics~(\ref{kerrex}). The metric~(\ref{kerrex}) with $\tilde{m}$ in ~(\ref{massBH}) and $\Delta$ in ~(\ref{Delta2}), is a prototype non-Kerr black hole with additional parameter $\ell_0$ due to hair, and deviation parameter $\alpha$. The metric~(\ref{kerrex}) is mathematically the same as the Kerr black  hole \cite{Kerr:1963ud} but the mass $M$ is replaced by $\tilde{m}(r)$. 

\begin{figure}[t]
	\begin{centering}
		\begin{tabular}{ccc}
		    \includegraphics[scale=0.75]{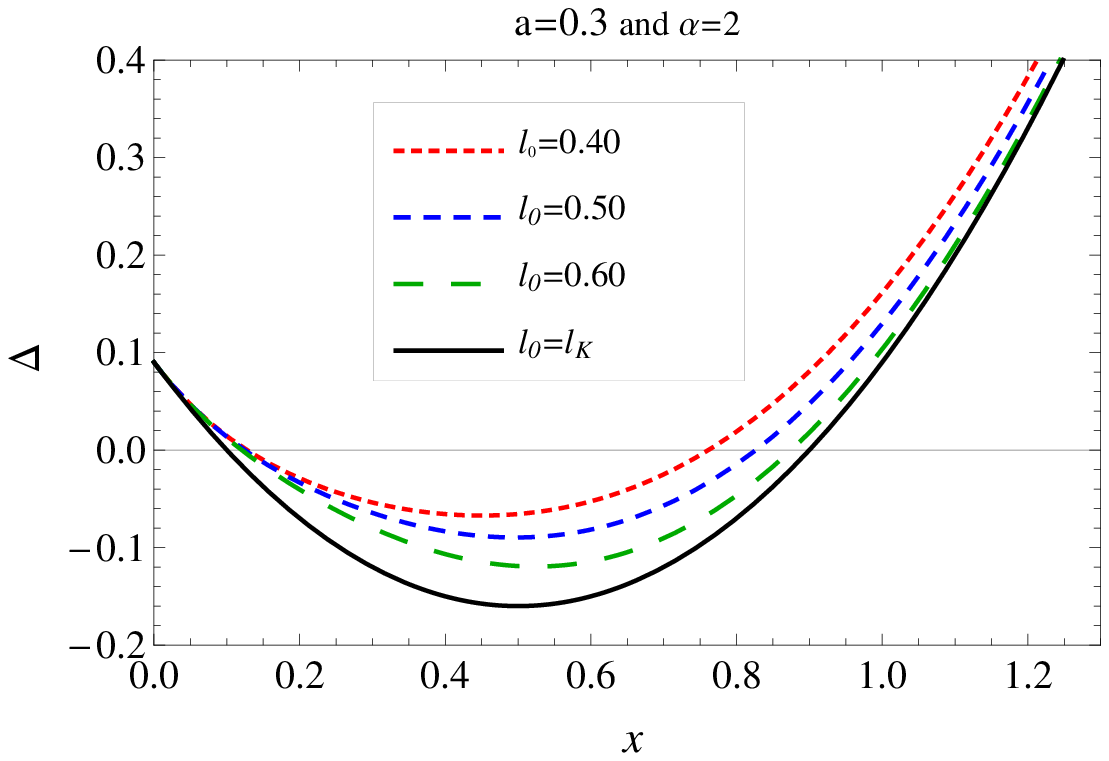}&
		    \includegraphics[scale=0.75]{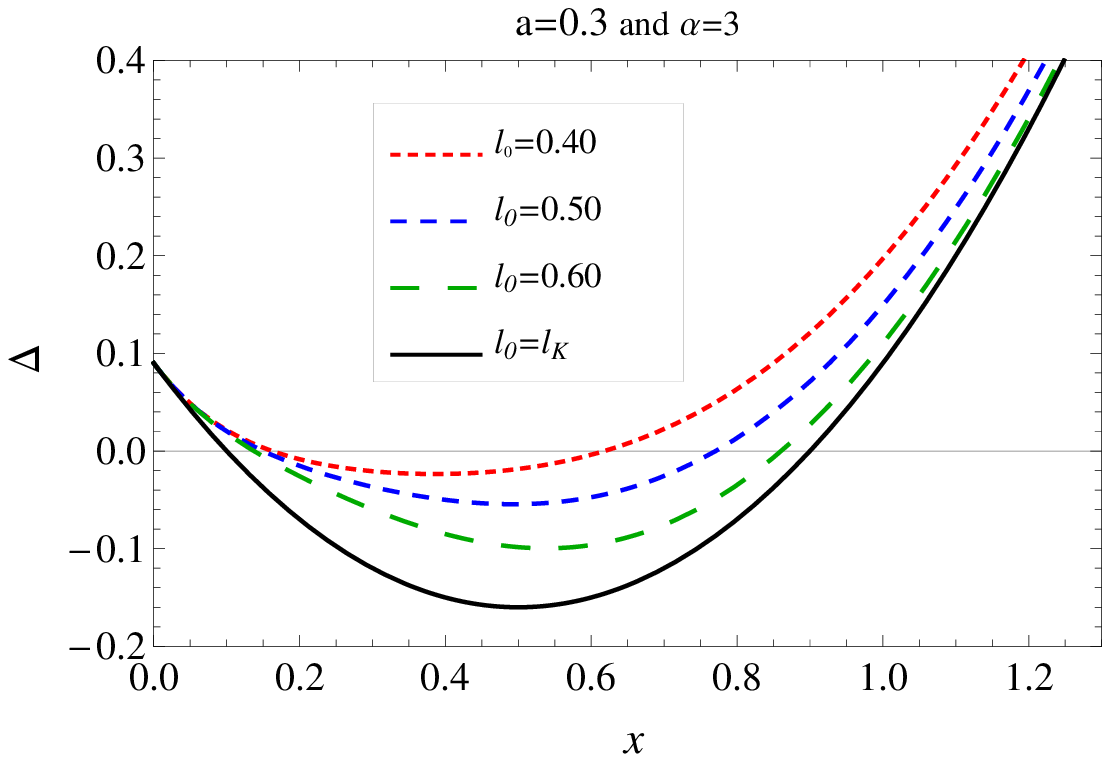}\\
		    \includegraphics[scale=0.75]{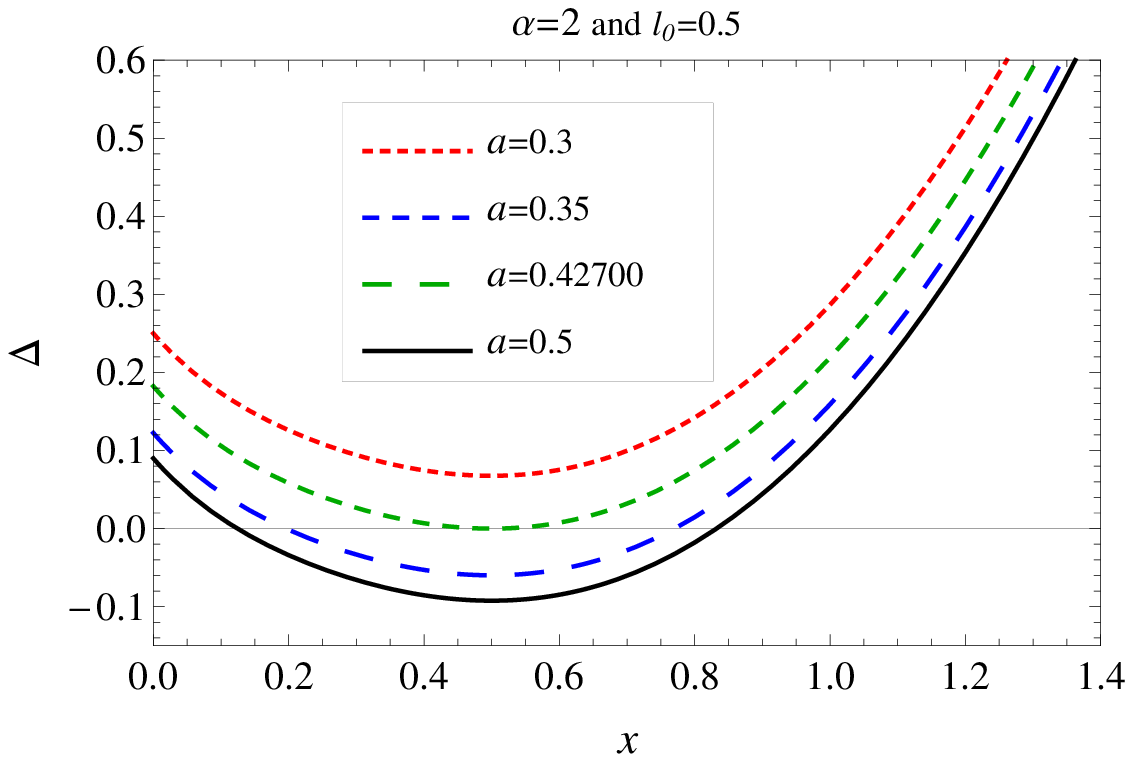}&
		    \includegraphics[scale=0.75]{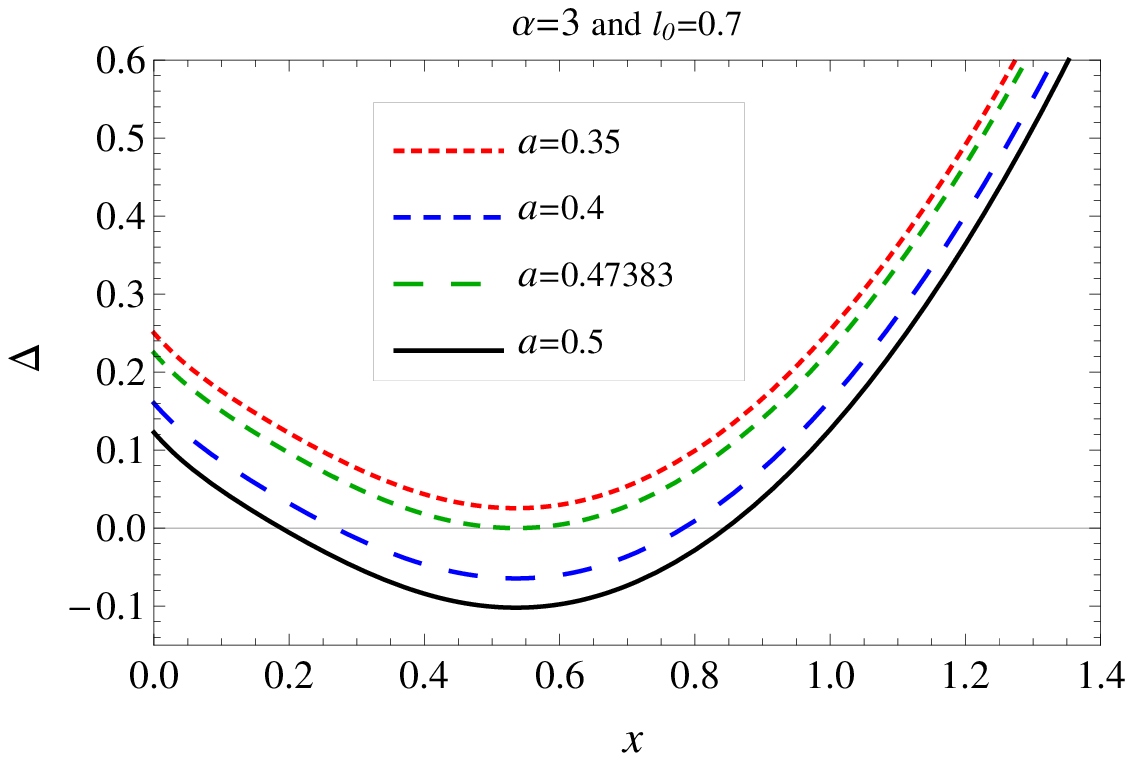}
		 \end{tabular}
	\end{centering}
	\caption{The behavior of the horizon ($\Delta$ vs $x$) of hairy black holes. The case  $\ell_{0}=\ell_{K}$ (solid line) corresponds to the Kerr black hole (top).}\label{plot1}		
\end{figure}

\begin{figure}[t]
	\begin{centering}
		\begin{tabular}{ccc}
		    \includegraphics[scale=0.75]{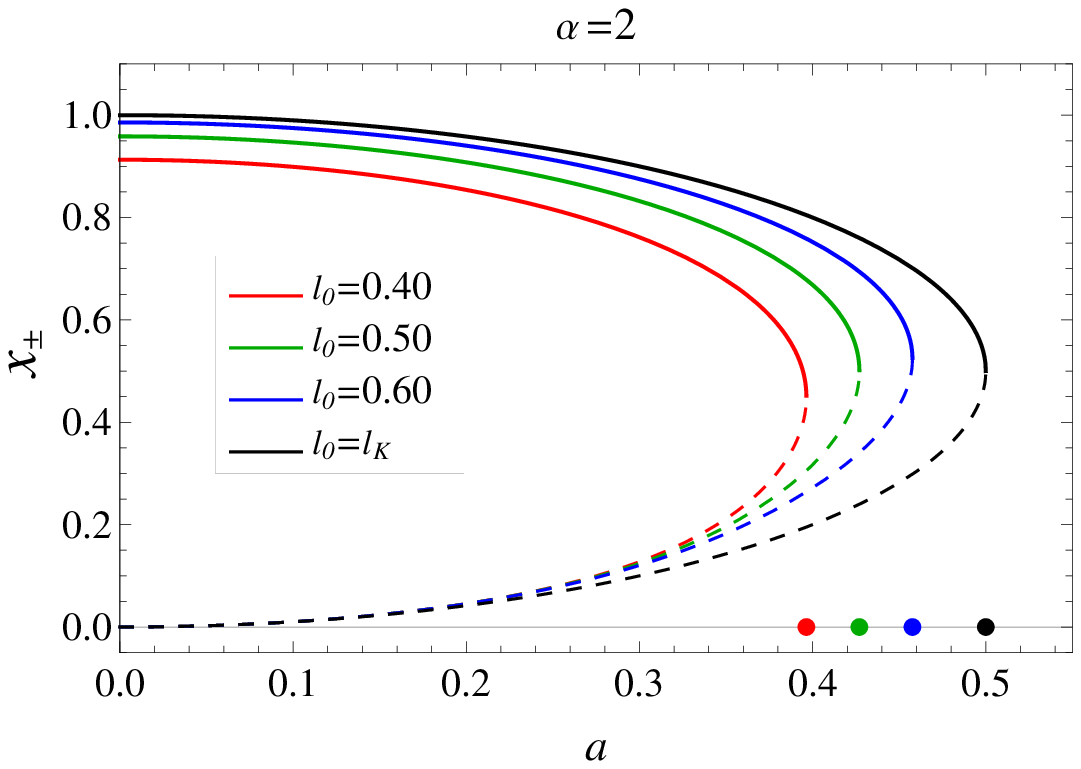}&
		    \includegraphics[scale=0.75]{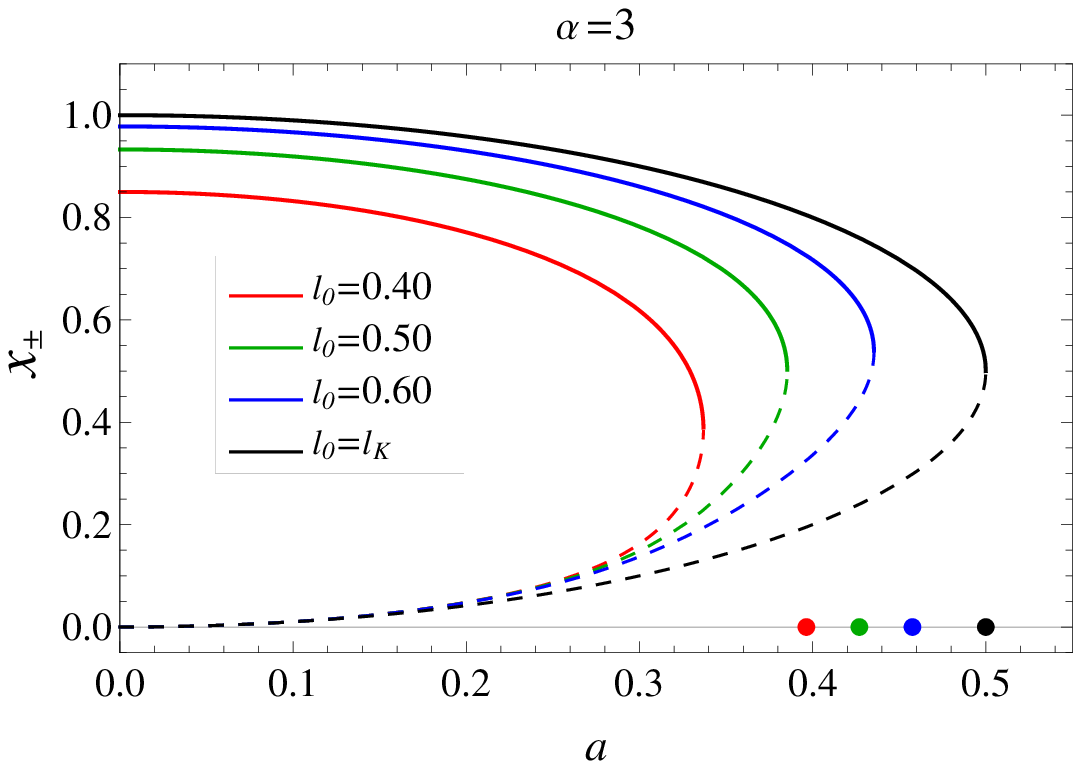}\\
		    \end{tabular}
	\end{centering}
	\caption{The event horizon (solid lines) and the Cauchy horizon (dotted lines) are shown  for the hairy Kerr black holes in comparison with the Kerr black hole ($\ell_{0}=\ell_{K}$). Points on the horizontal axis correspond to extremal values of $a$.}\label{plot2}		
\end{figure}  

\begin{figure}
	\begin{centering}
		\begin{tabular}{ccc}
		    \includegraphics[scale=0.75]{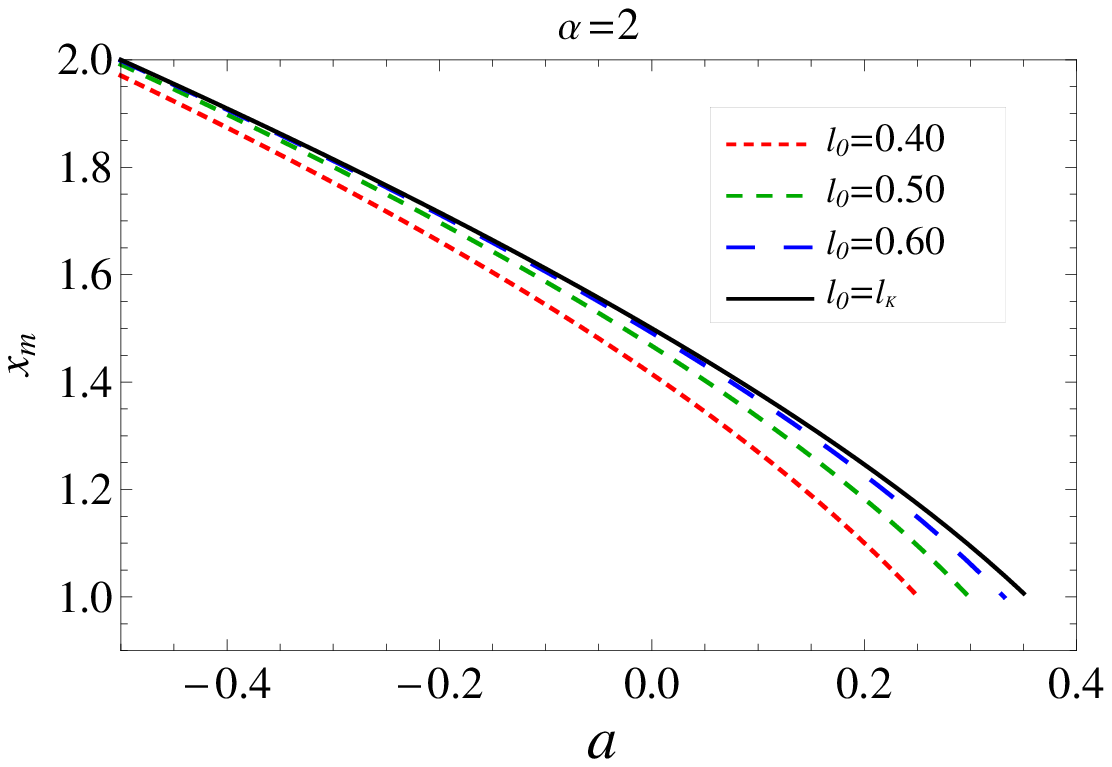}&
		    \includegraphics[scale=0.75]{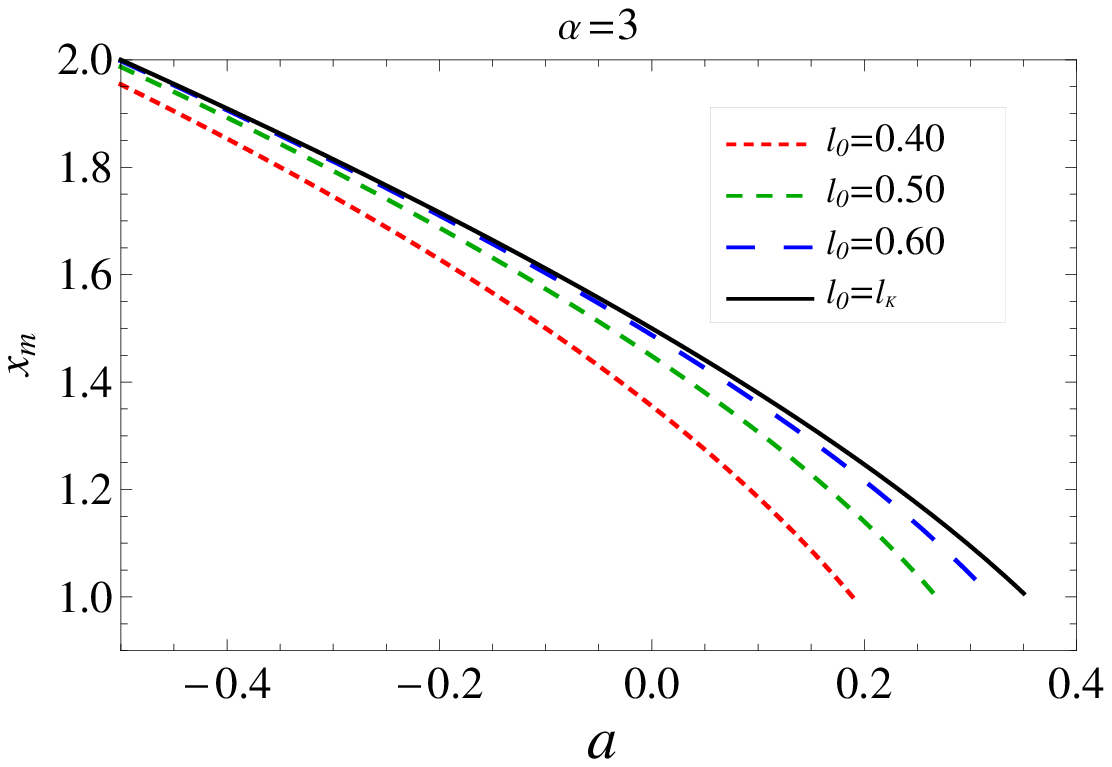}\\
		  \end{tabular}
	\end{centering}
	\caption{The  radii of photon spheres ($x_m$ vs $a$)  for the hairy black holes (dashed line) in comparison with the Kerr black hole (solid line) ($\ell_{0}=\ell_{K}$).}\label{plot3}		
\end{figure}

\section{Strong gravitational lensing by  rotating black holes}\label{sgl}
In this section, we will study the gravitational lensing by the hairy Kerr black holes~(\ref{kerrex})  to investigate  how the parameters $\alpha$ and $\ell_0$ affect  the lensing observables in the strong field limit. The strong gravitational lensing is determined by the lens equation and the equation for the deflection angle.  Henceforth, we  measure all quantities $r, a, t$,  and $l_0$ in units of $2M$ and, i.e., we set $M=1/2$ and use $x$ instead of radius $r$ to obtain
\begin{eqnarray}\label{metric2}
ds^2 & = & - \left[ 1- \frac{x \, \tilde{m}(x)}{\Sigma} \right] dt^2 +
\frac{\Sigma}{\Delta}\,dx^2 + \Sigma\, d \theta^2- \frac{2\, a\, x\,  \tilde{m}(x)\, 
}{\Sigma  } \sin^{2}\theta \, dt \; d\phi \nonumber
\\ & + & \left[x^2+ a^2 +
 \frac{a^2 x\,  \tilde{m}(x)   }{\Sigma} \sin^2 \theta
\right] \sin^2 \theta \, d\phi^2,
\end{eqnarray}
where $\Sigma = x^2 + a^2 \cos^2\theta$, $\Delta=x^2 + a^2 - x\; \tilde{m}(x)$,  and $\tilde{m}(x) = 1 -\alpha \, x \, e^{-2 \, x/(1-\ell_0)}$. The metric (\ref{metric2}) has a singularity when $\Sigma \neq 0 $ and  $\Delta=0$ corresponding to the event horizon, which are  the  zeroes of $g^{rr}=\Delta=0$, i.e.,
\begin{equation} \label{horizoneq}
x^2+a^{2}-\,x\,\left[1-\alpha \, x \, e^{-2 \, x/(1-\ell_0)} \right]=~0.
\end{equation}
One can find that there exists nonzero values of $a$, $\alpha$, and $\ell_0$ for which Equation~(\ref{horizoneq}) admits two positive roots ($x_{\pm}$) corresponding to Cauchy $(x_{-})$ and event horizons $(x_{+})$, which  are depicted in Fig.~\ref{plot1}.  The parameter $\ell_0$ should be greater than the critical value $\ell_c$ for the existence of the horizon (cf. Fig.~\ref{plot1}) and $\ell_0=\ell_K$ correspond to the Kerr black hole. The $\ell_c$ occurs when $\Delta=0$ has two equal real roots which can be numerically calculated e.g., for $\alpha=3$, $a=0.3$ we have $\ell_c=0.1036$, $x_c=0.2011$. The radius of the event horizon for hairy Kerr black holes are  smaller than that of the Kerr black hole (cf. Fig.~{\ref{plot1}}). The horizon shifts to larger radii when $\ell_0$ increases, reaching a maximum value corresponding to the Kerr horizon for $\ell_0=\ell_{\rm K} = 1$.  The hairy Kerr black holes, when compared to the Kerr black hole, have a small extreme rotation parameter (cf. Fig.~\ref{plot2}) and thereby the spacetime structure changes in the strong field region.

Next, to discuss the strong deflection of light by hairy Kerr black holes  we shall consider light rays strictly in the equatorial plane ($\theta=\pi/2$) and the metric  (\ref{metric2}) simplifies to
\begin{eqnarray}\label{NSR}
\mathrm{ds^2}=-A(x)\,dt^2+B(x)\,dx^2 +C(x)\,d\phi^2-D(x)dt\,d\phi,
\end{eqnarray}
where
\begin{eqnarray}
A(x)&=& 1-\frac{1-\alpha \, x \, e^{-2 \, x/(1-\ell_0)}}{x},\;\;\;\;\;~~~~
B(x)=\frac{x^2}{\Delta},
\nonumber\\
C(x)&=& \frac{\left(x^{2}+a^{2}\right)^{2}
-a^{2}\, \Delta }{x^2},
\;\;\;~~~~~~~~
D(x)=\frac{2\,a\left[1-\alpha \, x \, e^{-2 \, x/(1-\ell_0)} \right]}{x}.
\end{eqnarray}
The black hole metric (\ref{NSR}) admits two linearly independent killing vectors, 
$\eta^{\mu}_{(t)}=\delta^{\mu}_t $ and $\eta^{\mu}_{(\phi)}=\delta^{\mu}_{\phi}$, associated with the time translation and rotational invariance  \cite{Chander:1992pc}. The photon's trajectory is determined by two conserved quantities admitting the Killing vectors, i.e., the angular momentum $L$ and the total energy $E$. The null geodesic equations for the metric can be derived using the Hamilton-Jacobi method, and we get the following differential systems:
\begin{eqnarray}
\dot{t} &=& \frac{4C-2 L D}{4AC + D^2},\label{tdot}\\
\dot{\phi} &=& \frac{2D+4 A L}{4AC + D^2},\\
\dot{x} &=& \pm 2 \sqrt{\frac{C-DL-AL^2}{B(4AC + D^2)}},\label{xdot} \end{eqnarray}
where the dot indicates the derivative with respect to the affine parameter. Using Equation~(\ref{xdot}), the effective potential $V_{\text{eff}}$ for radial motion can be obtained as
\begin{eqnarray}\label{effpot}
V_{\text{eff}} &=& \frac{4(C-DL-AL^2)}{B(4AC + D^2)},
\end{eqnarray}
which characterizes the different types of possible orbits. In the asymptotic limit, depending on the effective potential, a light ray from the source at infinity approaches the black hole and may turn at some radius  $x_0$, only to escape towards the observer at infinity. It is a well-known fact that the deflection angle diverges as light approaches the photon sphere, and as such, there can be an infinite number of images just outside the photon sphere. 
The conditions for the unstable  photon sphere are \cite{Harko:2009xf}
\begin{eqnarray}
V_{\text{eff}}=\frac{dV_{\text{eff}}}{dx}\Big|_{(x_0=x_m)}=0, ~~~~~~~~\frac{d^2 V_{\text{eff}}}{dx^2}\Big|_{(x_0=x_m)}<0,
\end{eqnarray}
where $x_m$ is the radius of the photon sphere and $x_0$ is the distance of the light's minimum approach towards the black hole.
The photon sphere radius is given the equation
\begin{eqnarray}\label{ps}
A(x)C'(x)-A'(x)C(x) + L(A'(x)D(x) - A(x)D'(x)) = 0.  
\end{eqnarray}

The photon orbit radius, $x_m$, is the largest root of  Equation~(\ref{ps}). Thus, circular photon orbits simultaneously  satisfy $\dot{x}=\ddot{x}=0$ and the photon sphere which constitutes the unstable photon orbits additionally satisfies $\dddot{x}>0$. It turns out that the photon sphere depends on the hair ($\ell_0$), deformation parameter ($\alpha$), and the rotation parameter ($a$) (cf. Fig.~\ref{plot3}). The Fig.~\ref{plot3} shows  the decrease of radius $x_m$ with the rotation parameter and the $x_m$ of hairy black holes is smaller than the Kerr black hole. The photons are allowed  to get closer to the black hole for positive $a$.  The impact parameter, which is the perpendicular distance from the center of mass of the lens to the tangent of the null geodesics and which remains constant throughout the trajectory, coincides with the angular momentum $L$ in the equatorial plane.  The turning point is marked  by $\dot{x} = 0$ and  effective potential vanishes $V_{\text{eff}} =0$,   implying 
\begin{eqnarray}\label{angmom}
L &=& u(x_0) = \frac{a P(x_0)+x_0 \sqrt{a^2+x_0[x_0+ P(x_0)]}}{x_0+P(x_0)}\\ 
P(x_0) &=& \alpha x_0e^{2x_0/(1-\ell_0)}-1.
\end{eqnarray}
The subscript $0$ is defined by  $A(x_0)=A(x)$. Equation~(\ref{angmom}) relates the impact parameter $u$  and minimum distance $x_0$.  Photons winding in the same sense (prograde or direct  photons) as that of black hole rotation form different orbits than those winding in the opposite direction (retrograde photons). We fix the counterclockwise winding of light rays by choosing the positive sign before the square bracket in Equation~(\ref{angmom}). For $a>0$, the black hole also rotates in the counterclockwise direction, while for $a<0$, the black rotates in the opposite direction of photon winding. 

The deflection angle in rotating stationary 
spacetime described by the line element (\ref{NSR}), at  the closest distance approach $x_0$, is given by \cite{Bozza:2002zj}
\begin{eqnarray}\label{bending1}
\alpha_{D}(x_0)=I(x_0)-\pi,
\end{eqnarray} 
where
\begin{eqnarray}\label{bending2}
I(x_0) = 2 \int_{x_0}^{\infty}\frac{d\phi}{dx} dx
= 2\int_{x_0}^{\infty}\frac{\sqrt{A_0 B }\left(2AL+ D\right)}{
\sqrt{4AC+D^2}\sqrt{A_0 C-A C_0+L\left(AD_0-A_0D\right)}} dx,
\end{eqnarray}

\begin{figure}[t]
	\begin{centering}
		\begin{tabular}{ccc}
		    \includegraphics[scale=0.75]{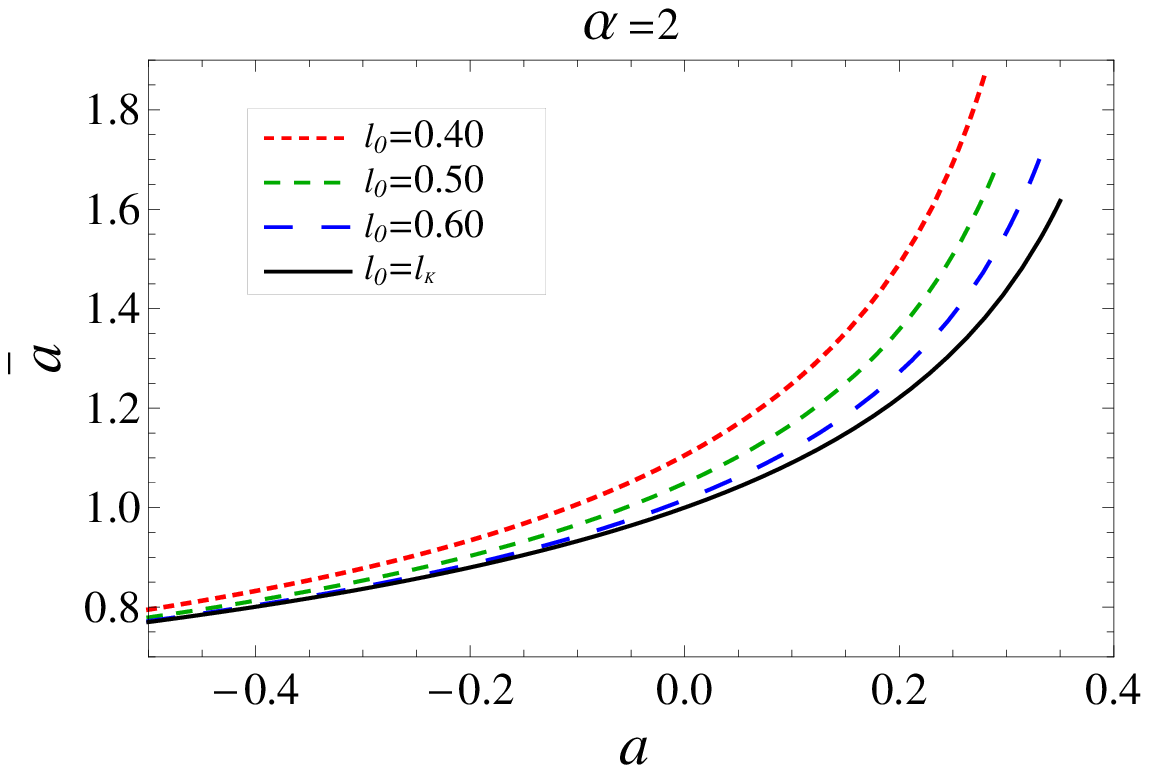}&
		    \includegraphics[scale=0.75]{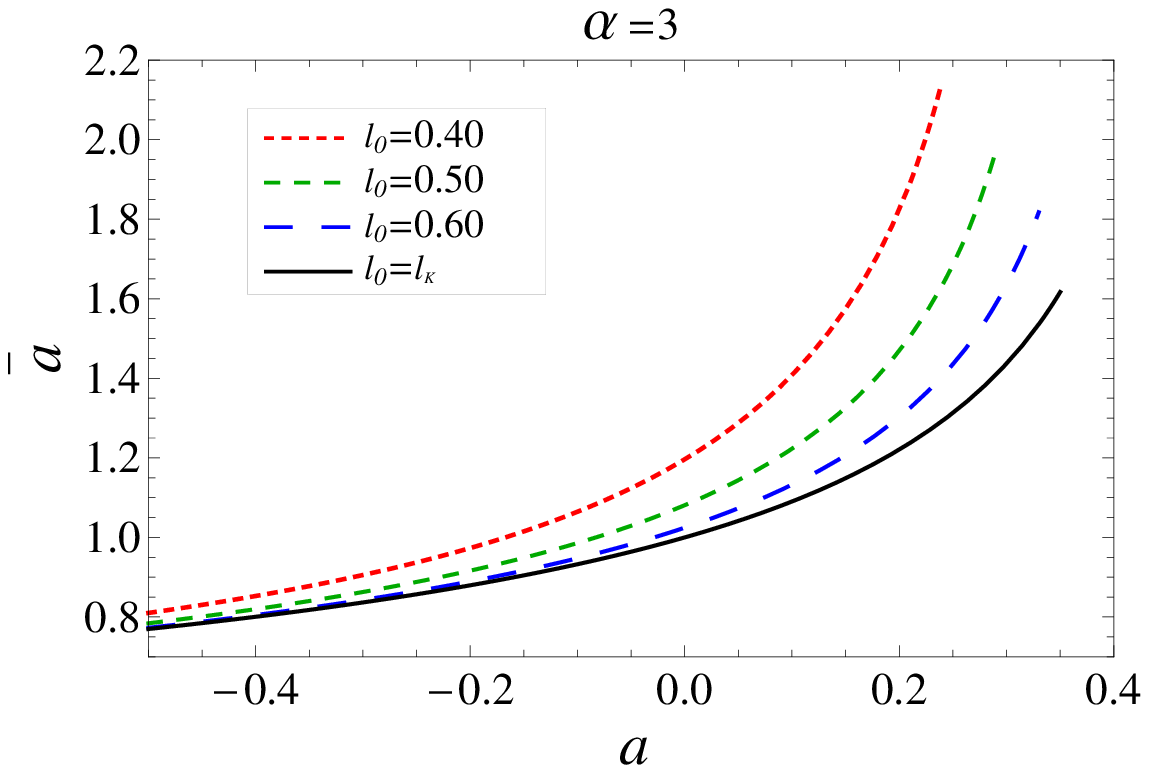}\\
		    \includegraphics[scale=0.75]{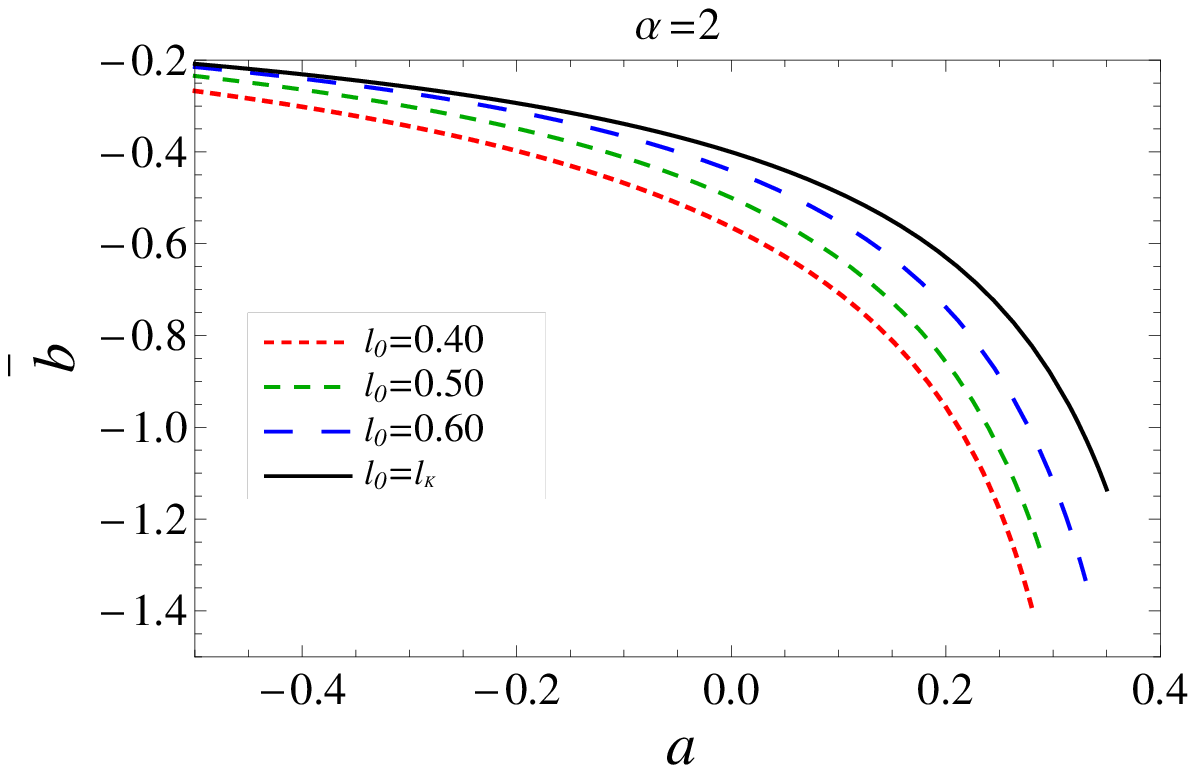}&
		    \includegraphics[scale=0.75]{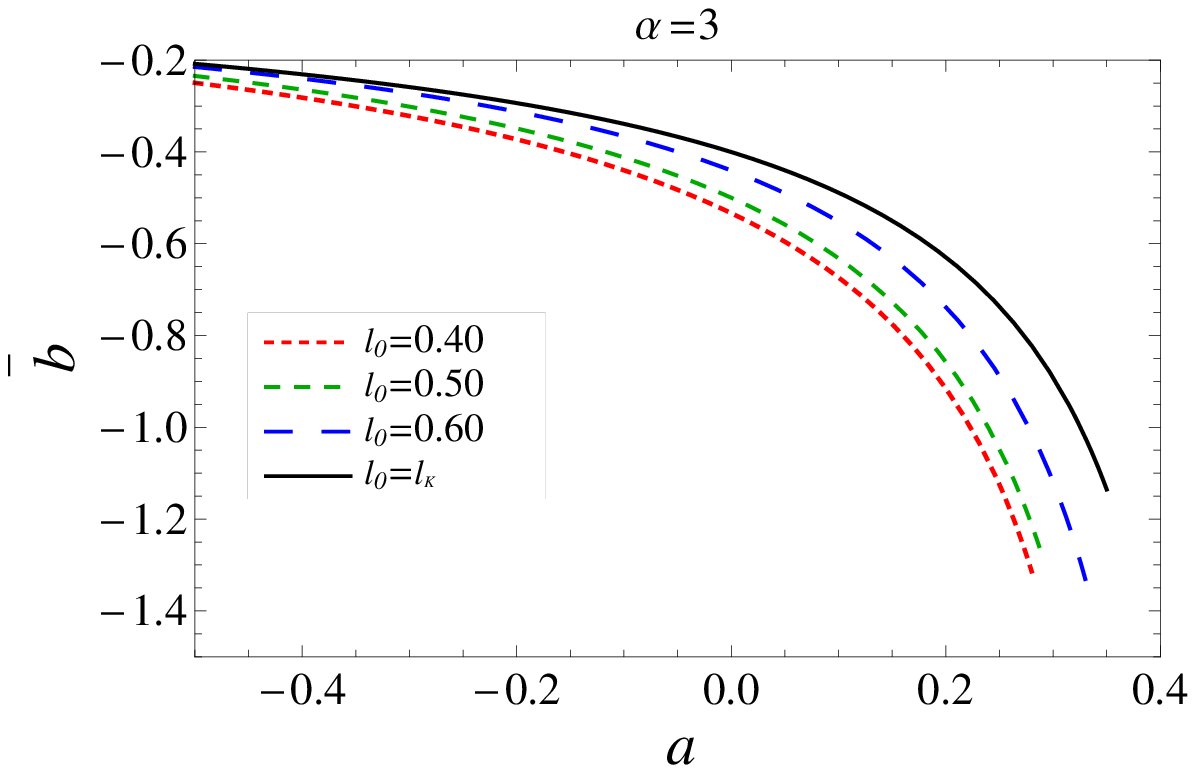}\\
		      \includegraphics[scale=0.75]{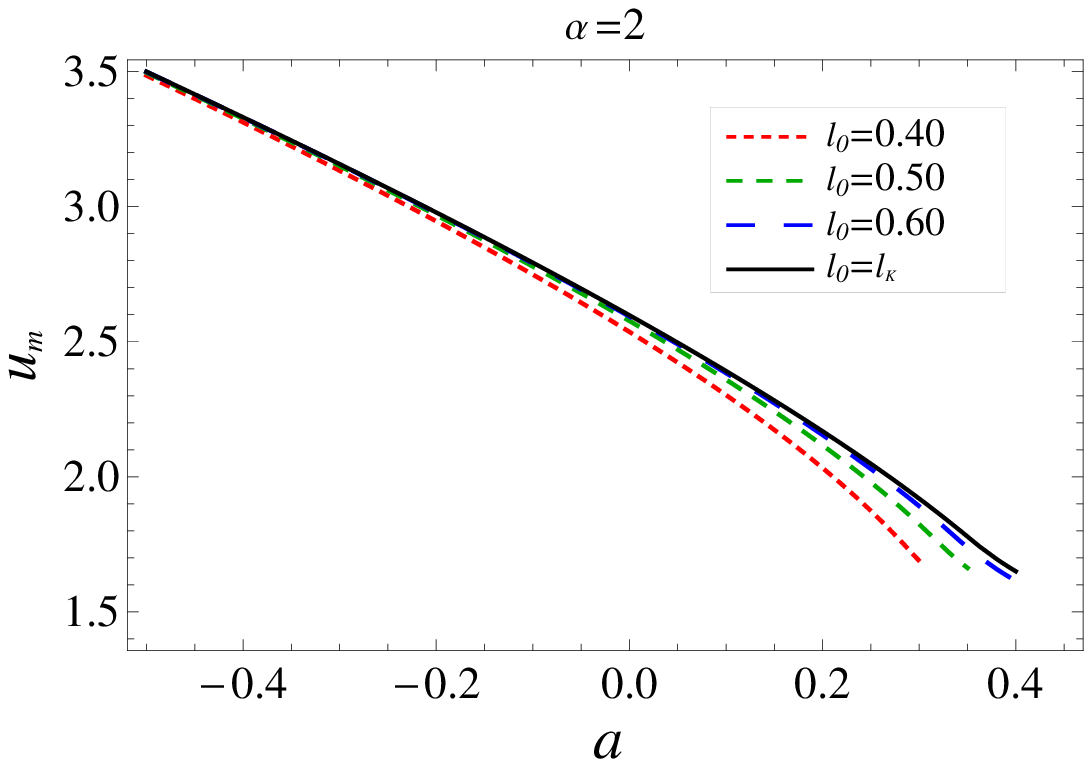}&
		    \includegraphics[scale=0.75]{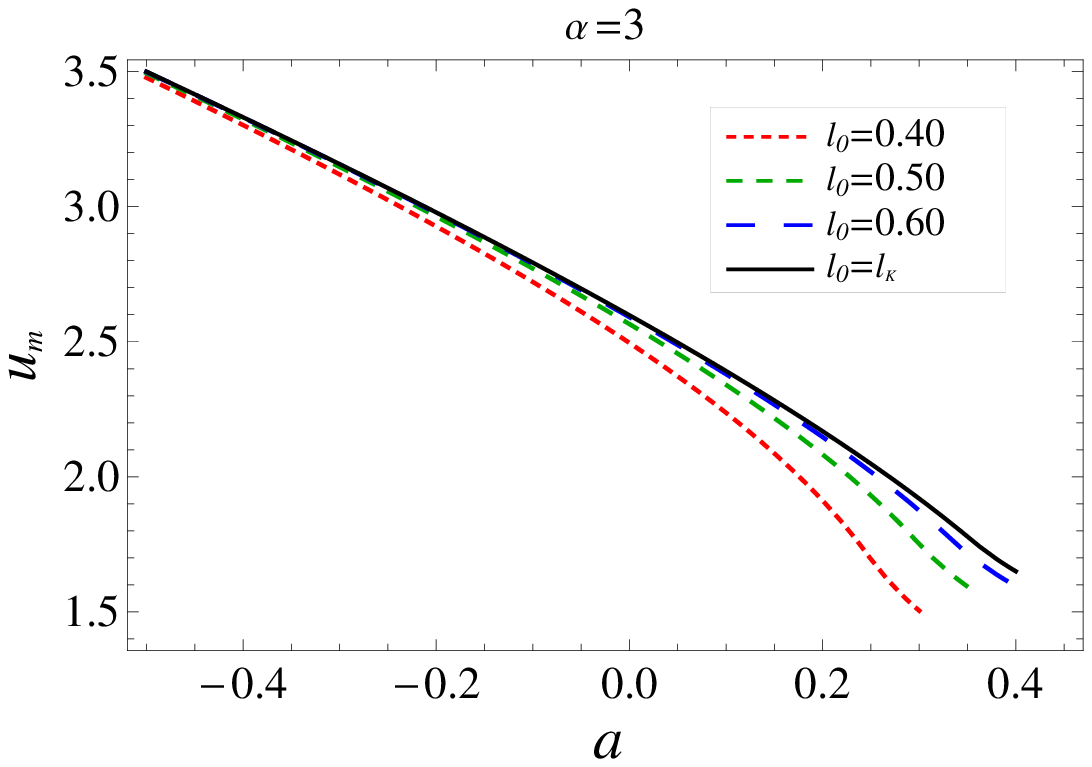}
		 \end{tabular}
	\end{centering}
	\caption{The behavior of  lensing coefficients $\bar{a}$, $\bar{b}$, and $u_m$ is shown as a function of black hole spin  $a$ for the hairy Kerr black holes (dashed lines) in comparison with the Kerr black hole 
	($\ell_{0}=\ell_{K}$) (solid lines).}\label{plot4}		
\end{figure}  

However, the integral in Equation~(\ref{bending2}) can not be solved in an explicit form. The deflection angle is very small in the weak deflection limit (WDL), and an approximate solution can be obtained. But the classical WDL is invalid when dealing with lensing in a strong gravitational field. For solving this problem one could seek a particular function to replace the integral as done in \cite{Darwin:1959,Eiroa:2004gh}, but a much more effective way to handle the integral in Equation~(\ref{bending2}) is to expand the deflection angle in the strong deflection limit (SDL) near the photon sphere  \cite{Bozza:2002zj,Tsukamoto:2016jzh}. This method provides an analytical representation of the deflection angle and a  straightforward and efficient connection between the coefficients and observables.

Next, we introduce the variable  $z=1 - x_0 / x$ \cite{Ghosh:2020spb}, and rewrite the integral (\ref{bending2}) as
\begin{eqnarray}\label{integral}
I(x_0)=\int_{0}^{1} R(z,x_0)f(z,x_0)dz,
\end{eqnarray}
where
\begin{eqnarray}
R(z,x_0)=\frac{2x^2}{x_0} \frac{\sqrt{B}\left(2A_0AL+A_0D\right)}{\sqrt{CA_0}\sqrt{4AC+D^2}},
\end{eqnarray}
\begin{eqnarray}\label{fz}
f(z,x_0)= \frac{1}{\sqrt{A_0-A\frac{C_0}{C}+\frac{L}{C}\left(AD_0-A_0D\right)}}.
\end{eqnarray}
 The function $R(z,x_0)$ is a regular for all values of $z$ and $x_0$, but  $f(z,x_0)$ diverges when $z \to 0$. Thus, the deflection angle becomes unbounded near $z=0$, where the photon approaches circular photon orbit radius. The function $f(z,x_0)$ can be approximated as 
\begin{eqnarray}
f(z,x_0)\sim f_0(z,x_0)=\frac{1}{\sqrt{c_1 z + c_2 z^2}},
\end{eqnarray}

\begin{table}[!htp]
\centering
\setlength{\tabcolsep}{0pt}
\begin{tabular*}{\textwidth}{@{\extracolsep{\fill}\quad}lccc} 
\hline\hline\\
{$a$ } &  {$\bar{a}$} & { {$\bar{b}$}} &$u_m/R_s$ 
\\ \hline
\hline
\\
-0.2 & 0.879625 & -0.293619 & 2.97879 \\
\hline \\ 
-0.1 & 0.932575 & -0.338872 & 2.79283 \\
\hline \\
0.0 & 1.& -0.40023 & 2.59808  \\            
\hline \\
0.1    & 1.0903 & -0.48879 & 2.39162 \\  
\hline \\
0.2   & 1.22095 & -0.629289 & 2.16856 \\  
\hline\hline
\end{tabular*}
\caption{The lensing coefficients for the Kerr black hole ($\alpha=0$)  compared with  Schwarzschild  black holes ($a=0$).
\label{table01} }
\end{table}  

\begin{table}[!htp]
\centering
\setlength{\tabcolsep}{0pt}
\begin{tabular*}{\textwidth}{@{\extracolsep{\fill}\quad}lcccc} 
\hline\hline\\
{$a$ } & {$\ell_0$} & {$\bar{a}$} & { {$\bar{b}$}} &$u_m/R_s$ 
\\ \hline
\hline
\\
\multirow{3}{*}{-0.2}& 0.40 &0.934129 & -0.397679 & 2.94631 \\
&0.50 &0.902917 & -0.348927 & 2.96902 \\
&0.60 &0.88599 & -0.312242 & 2.97708 \\
\hline \\
\multirow{3}{*}{-0.1} &0.40 &1.00671 & -0.467797 & 2.74872 \\
&0.50 &0.965757 & -0.411963 & 2.77877 \\
&0.60 &0.94241 & -0.36584 & 2.79013 \\
\hline \\
\multirow{3}{*}{0.0} & 0.40 & 1.10507 & -0.564093 & 2.53661  \\    
& 0.50 & 1.04924 & -0.499821 & 2.57728 \\  
& 0.60 & 1.01582 & -0.44049 & 2.59366  \\  
\hline \\
\multirow{3}{*}{0.1}   & 0.40 & 1.24907 & -0.707259 & 2.30279 \\  
& 0.50 & 1.16807 & -0.631789 & 2.35967   \\   
& 0.60 & 1.11732 & -0.551929 & 2.3841  \\     
\hline \\
\multirow{3}{*}{0.2}   & 0.40 & 1.49006 & -0.955554 & 2.03255 \\  
& 0.50 & 1.35766 & -0.856922 & 2.1165   \\   
& 0.60 & 1.27192 & -0.737659 & 2.1549  \\         
  \hline\hline
\end{tabular*}
\caption{The lensing coefficients for the hairy Kerr black holes ($\alpha=2$)   compared with the hairy Schwarzschild  black holes ($a=0$).
\label{table02} }
\end{table}

\begin{table}[!htp]
\centering
\setlength{\tabcolsep}{0pt}
\begin{tabular*}{\textwidth}{@{\extracolsep{\fill}\quad}lcccc} 
\hline\hline\\
{$a$ } & {$\ell_0$} & {$\bar{a}$} & { {$\bar{b}$}} &$u_m/R_s$ 
\\ \hline
\hline
\\
\multirow{3}{*}{-0.2}

&0.40 & 0.973276 & -0.48179 & 2.92713 \\
&0.50 & 0.916446 & -0.3826 & 2.96381 \\
&0.60 & 0.889303 & -0.322049 & 2.97622 \\
\hline \\

\multirow{3}{*}{-0.1}
&0.40 & 1.06455 & -0.582964 & 2.72137 \\
&0.50 & 0.9859 & -0.459058 & 2.7711 \\
&0.60 & 0.947615 & -0.380349 & 2.78875 \\
\hline \\

\multirow{3}{*}{0.0} & 0.40 & 1.19637 & -0.728376 & 2.49574  \\ 
& 0.50 & 1.08101 & -0.569035 & 2.56555 \\  
& 0.60 & 1.0244 & -0.462847 & 2.59138  \\  
\hline \\
\multirow{3}{*}{0.1}   & 0.40 & 1.40842 & -0.953293 & 2.23707 \\  
& 0.50 &1.22295 & -0.741972 & 2.3407   \\   
& 0.60 & 1.13255 & -0.588726 & 2.38013  \\     
\hline \\
\multirow{3}{*}{0.2}   & 0.40 & 1.82412 & -1.3549 & 1.91206 \\  
& 0.50 & 1.4691 & -1.06068 & 2.08267 \\   
& 0.60 & 1.30263 & -0.805935 & 2.14742  \\         
\hline\hline
\end{tabular*}
\caption{The lensing coefficients for the hairy Kerr black holes ($\alpha=3$)  compared with  the hairy Schwarzschild  black holes ($a=0$).
\label{table03} }
\end{table}  

where $c_1$ and $c_2$ are,  respectively,  obtained by Taylor expansion of the argument of  square root in $f(z,x_0)$. By assuming the closest approach distance $x_0$ to not be  too much larger than $x_m$, the deflection angle can be written  as \cite{Weinberg:1972,Bozza:2002zj} 

\begin{eqnarray}\label{def}
\alpha_{D}(\theta)=-\bar{a} \log\Big(\frac{\theta D_{OL}}{u_m}-1\Big)+ \bar{b} + \mathcal{O}\left(u-u_m\right),
\end{eqnarray}
where $u\approx \theta D_{OL}$ is the impact parameter and  $ D_{OL}$ is the distance between the observer and the lens. The strong deflection  coefficients $\bar{a}$ and $\bar{b}$ of the strong field limit, respectively, read 
\begin{eqnarray}\label{abar}
\bar{a} = \frac{R(0,x_m)}{2\sqrt{{c_2}_m}}, ~~~ \textrm{and}~~~ \bar{b} = -\pi +I_R(x_m) + \bar{a} \log\frac{c x_m^2 }{u_m^2}\\
I_R(x_m) = \int_{0}^{1} [R(z,x_m)f(z,x_m)-R(0,x_m)f_0(z,x_m)]dz,
\end{eqnarray}
\begin{equation}\label{um}
u-u_m = c(x_0-x_m)^2    
\end{equation}
\begin{figure}[t]
	\begin{centering}
		\begin{tabular}{ccc}
		    \includegraphics[scale=0.7]{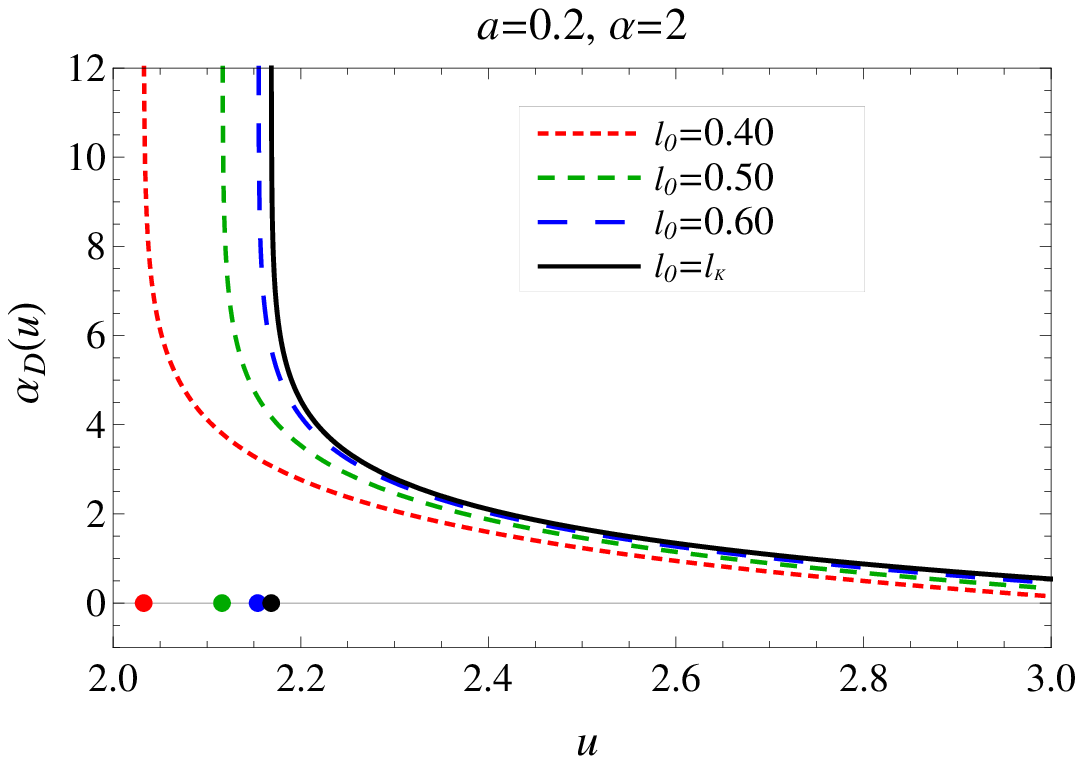}&
		    \includegraphics[scale=0.7]{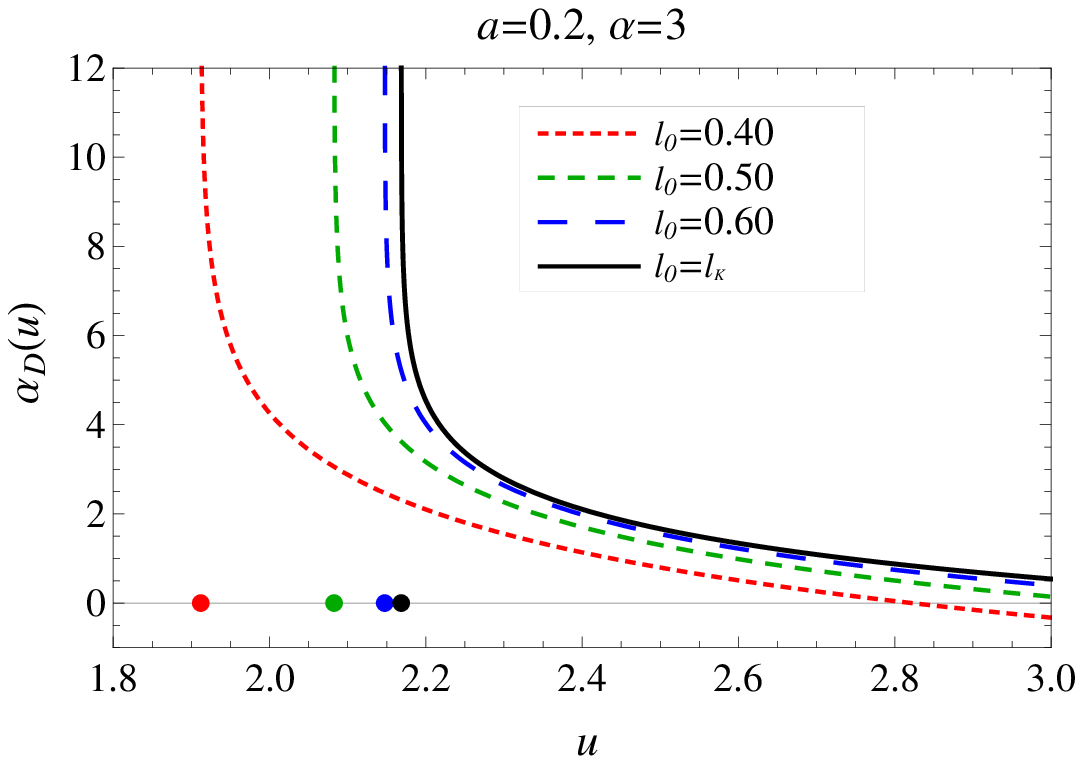}\\
		     \includegraphics[scale=0.7]{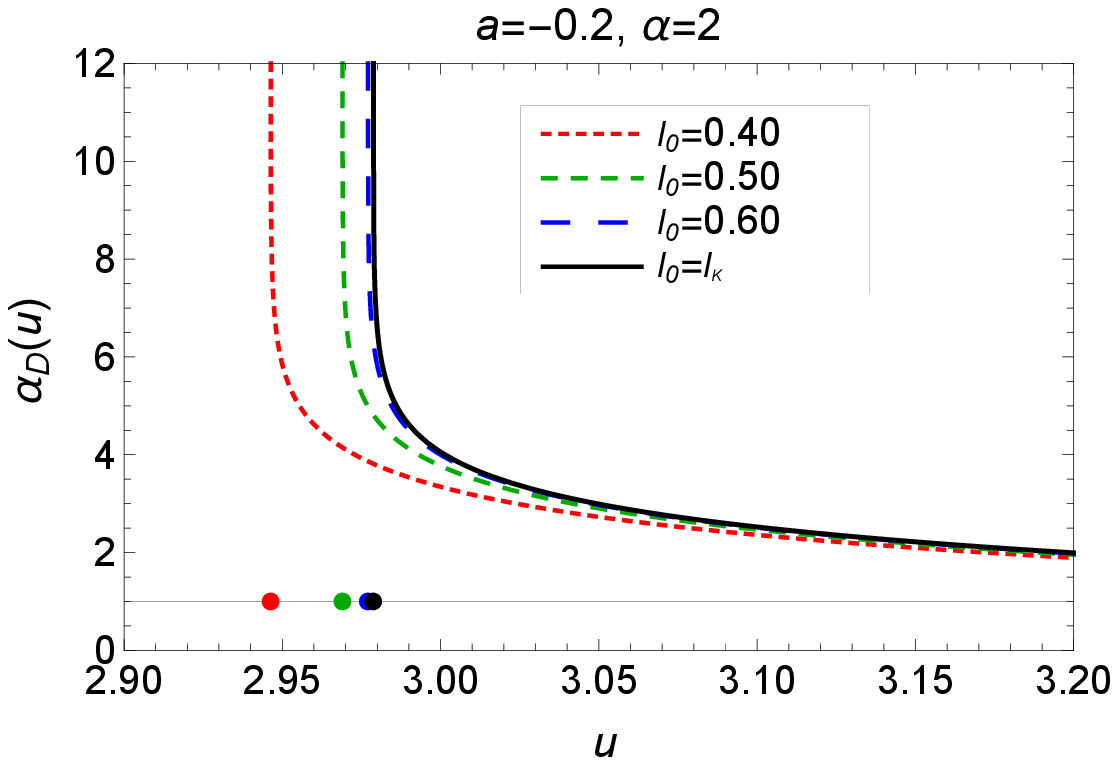}&
		    \includegraphics[scale=0.7]{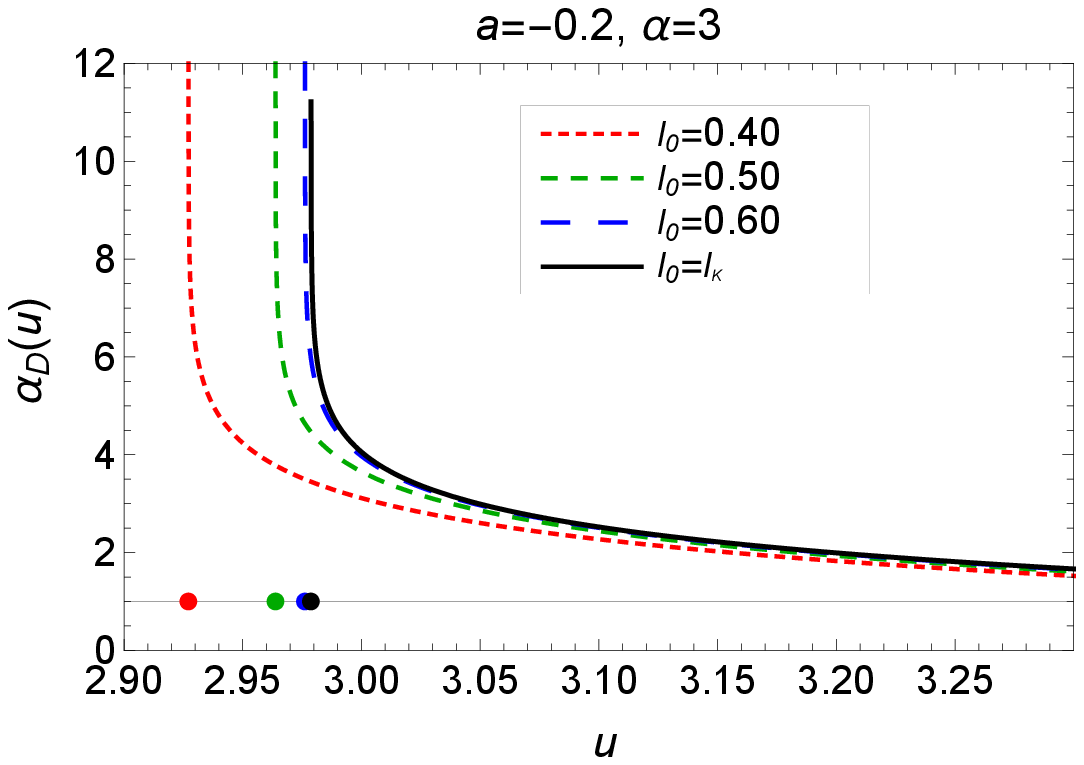}\\
		    \end{tabular}
	\end{centering}
	\caption{The variation of the deflection angle  as a function of impact parameter $u$ is shown for the hairy black holes (dashed lines) in comparison with the Kerr black ($\ell_{0}=\ell_{K}$) (solid line). Points on the horizontal axis represent the values of the impact parameter $u=u_m$ at which the deflection angle diverges.}\label{plot5}		
\end{figure}  

Using Eqs.~(\ref{def})--(\ref{um}), we can study the behavior of strong field lensing in hairy Kerr black holes. The  strong field deflection coefficients, $\bar{a}$, $\bar{b}$, and $u_m$,  are plotted against the angular spin in Fig.~\ref{plot4}, which shows that  $\bar{a}$ and $\bar{b}$ increase and decrease, respectively, with $a$. The minimum impact parameter $u_m$ decreases in a similar fashion to $x_m$. For the retrograde  photons, the lensing coefficients are very close to the Kerr black hole.  The lensing coefficients $\bar{a}$ and $\bar{b}$ in hairy Kerr black holes  diverge at lower values of spin in comparison to standard Kerr black hole  (cf. Fig.~\ref{plot4} and  Table~{\ref{table01}--\ref{table03}}). For fixed values of parameters, the deflection angle diverges at a lower impact parameter for larger $\alpha$ (cf. Fig.~\ref{plot5}). Moreover, the values of $u_m$  get smaller for hairy Kerr black holes. It can also be inferred from Fig.~\ref{plot5} that strong lensing is valid only when the impact parameter is very close to $u_m$. We have also tabulated the deviation of lensing coefficients of hairy Kerr black holes with $\alpha=3$ from Kerr black holes in Table \ref{table04}.

\begin{table}[!htp]
\centering
\setlength{\tabcolsep}{0pt}
\begin{tabular*}{\textwidth}{@{\extracolsep{\fill}\quad}lcccc} 
\hline\hline\\
{$a$ } & {$\ell_0$} & {$\delta\bar{a}$} & { {$\delta\bar{b}$}} &$\delta u_m/R_s$ 
\\ \hline
\hline
\\

\multirow{3}{*}{-0.2}
&0.40& 0.0936514 & -0.188172 & -0.051656 \\
&0.50& 0.036821 & -0.0889811 & -0.0149769 \\
&0.60& 0.00967844 & -0.0284298 & -0.00256955 \\
\hline\\

\multirow{3}{*}{-0.1}
& 0.40 & 0.131973 & -0.244092 & -0.0714636 \\
& 0.50 & 0.0533249 & -0.120186 & -0.02173 \\
& 0.60 & 0.0150393 & -0.0414768 & -0.0040815 \\
\hline\\

\multirow{3}{*}{0.0} & 0.40 & 0.196367 & -0.328146 & -0.102341  \\ 
& 0.50 & 0.0810136 & -0.168805 & -0.0325222 \\  
& 0.60 & 0.0243984 & -0.0626169 & -0.00669681\\
\hline\\

\multirow{3}{*}{0.1}   & 0.40 & 0.318114 & -0.464503 & -0.154557 \\  
& 0.50 & 0.132646 & -0.253181 & -0.0509265   \\   
& 0.60 & 0.0422512 & -0.0999354 & -0.0114949  \\     
\hline \\

\multirow{3}{*}{0.2}   & 0.40 & 0.60317 & -0.725608 & -0.2565\\  
& 0.50 & 0.248156 & -0.431391 & -0.0858858 \\   
& 0.60 & 0.0816841 & -0.176647 & -0.0211412  \\         
\hline\hline
\end{tabular*}
\caption{Deviation  of the lensing coefficients of hairy black holes ($\alpha=3$) from the Kerr black hole where $\delta X = X_{\text{Kerr}}-X_{\text{hairy Kerr}}$.
\label{table04} }
\end{table}  

\section{Observables and relativistic images}\label{observables}
Let us assume that a light ray emitted by a source $S$ with angle $\beta$  is scattered by a black hole or lens L by deflection angle $\alpha_D$.  The source is seen as an image I at an angle $\theta$ by an observer. $\alpha_D$ is the total angle light and is deviated from its path by the  gravitational field of the lens while travelling from the source to the observer.  $D_{OL}$, $D_{LS}$, and $D_{OS}$ are the observer-lens, lens-source, and observer-source distances.  As long as the deflection angle is not vanishing, the closest approach distance $x_0$ is different than the impact parameter $u$.

In the lens equation, we consider the asymptotic approximation, i.e., both the source and the observer are not affected by the  curvature of the lens and lie in flat spacetime. It allows us to use Euclidean geometry relations to relate the various quantities and restore all the relativistic information in the deflection angle without losing generality. As discussed by the authors in
\cite{Bozza:2008ev}, the lens equation used in \cite{Oho:1987ev} would be adequate, without the need to resort to the exact lens equation. Accordingly, we introduce here a coordinate independent  lens equation \citep{Oho:1987ev} connecting the source and observer positions as
\begin{eqnarray}\label{oho}
\xi &=& \frac{D_{OL}+D_{LS}}{D_{LS}}\theta-\alpha_{D}(\theta),   
\end{eqnarray} 
where $\xi$ is the angle between the direction of source and optical axis as viewed from the lens.  The angle $\xi$ and $\beta$ are related via \cite{Bozza:2008ev}
\begin{eqnarray}\label{rel}
\frac{D_{OL}}{\sin(\xi-\beta)} &=& \frac{D_{LS}}{\sin \beta}.
\end{eqnarray}
We use the hypothesis of  small angles  $\alpha_D, \beta$, and $\theta$ in the classical lens equation so it makes sense to perform small angle approximation in all the trigonometric functions and reconsider  the exact lens equation. Also the lensing effects are  prominent when all the objects are almost aligned. Rewriting  Equation~(\ref{oho}) for small values of  $\beta$, $\xi$, and $\theta$, and  using Equation~(\ref{rel}), we get \cite{Bozza:2008ev}
\begin{eqnarray}\label{lensequation}
\beta &=& \theta -\frac{D_{LS}}{D_{OL}+D_{LS}} \alpha _D(\theta).
\end{eqnarray} 
It is interesting to see that for strong field lensing, where a ray of light emitted by the source $S$ follows multiple loops around the black hole before reaching the observer, a similar expression is obtained. However  $\alpha_{D}(\theta)$ is replaced by    $\alpha_{D}(\theta)-2n\pi = \Delta\alpha _n$ ,with  $n \in N $ and $ 0 < \Delta\alpha _n \ll 1 $ as 
 \begin{eqnarray}\label{lensequation1}
\beta &=& \theta -\frac{D_{LS}}{D_{OL}+D_{LS}} \Delta\alpha_n.
\end{eqnarray} 
Image positions are calculated using Equation~(\ref{lensequation}) for  given values of the angular position of source $\beta $ and the distances of observer and source from the black hole. As the photons approach the event horizon,  the deflection angle becomes greater than $2\pi$ such that at the critical impact parameter, it diverges.   Using Equations~(\ref{def}) and (\ref{lensequation1}), the angular separation between the optical axis and $n$-loop relativistic image can be written as a combination of two parts \cite{Bozza:2002zj},
\begin{eqnarray}\label{angpos}
\theta_n &=& \theta_n{^0} + \Delta\theta_n.
\end{eqnarray}
where 
\begin{eqnarray}
\theta_n{^0} &=& \frac{u_m}{D_{OL}}(1+e_n),\\
\Delta\theta_n &=& \frac{D_{OL}+D_{LS}}{D_{LS}}\frac{u_me_n}{D_{OL}\bar{a}}(\beta-\theta_n{^0}),\\
e_n &=& \text{exp}\left({\frac{\bar{b}}{\bar{a}}-\frac{2n\pi}{\bar{a}}}\right).
\end{eqnarray}
Here $\theta_n{^0}$ is the corresponding value of $\theta$ when $\alpha_D(\theta)=2n\pi$. $\Delta\theta_n$ is the correction term which is smaller than the main term $\theta_n{^0}$. The Equation~(\ref{angpos}) gives images only on the same side of the source. One can solve the same equation for $\beta<0$ to obtain the images on the other side. 
\begin{figure}[t]
	\begin{centering}
		\begin{tabular}{ccc}
		    \includegraphics[scale=0.8]{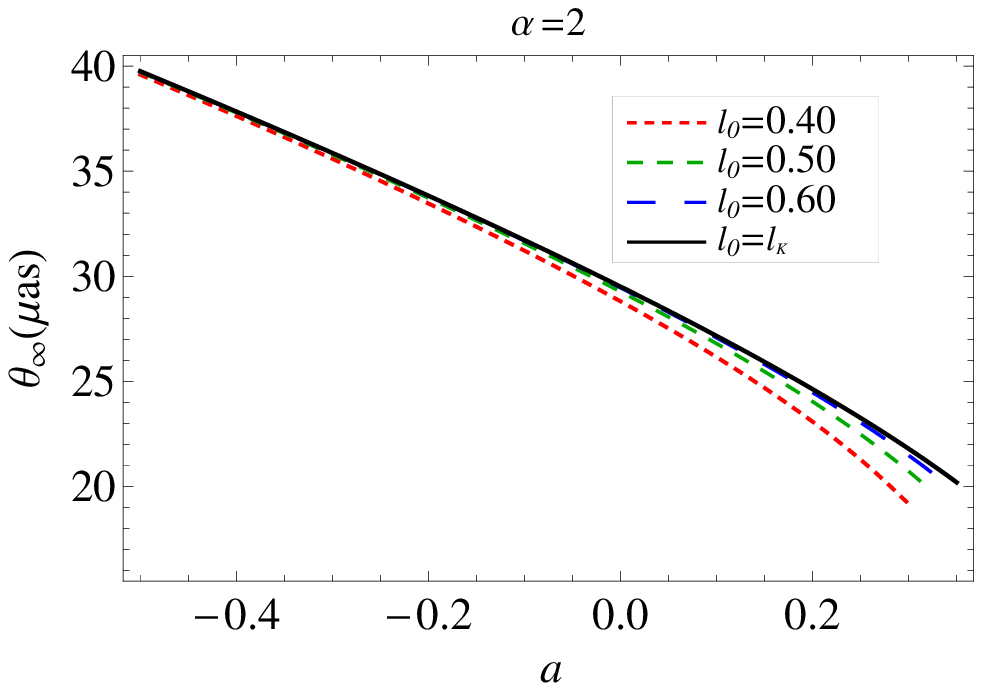}&
		    \includegraphics[scale=0.8]{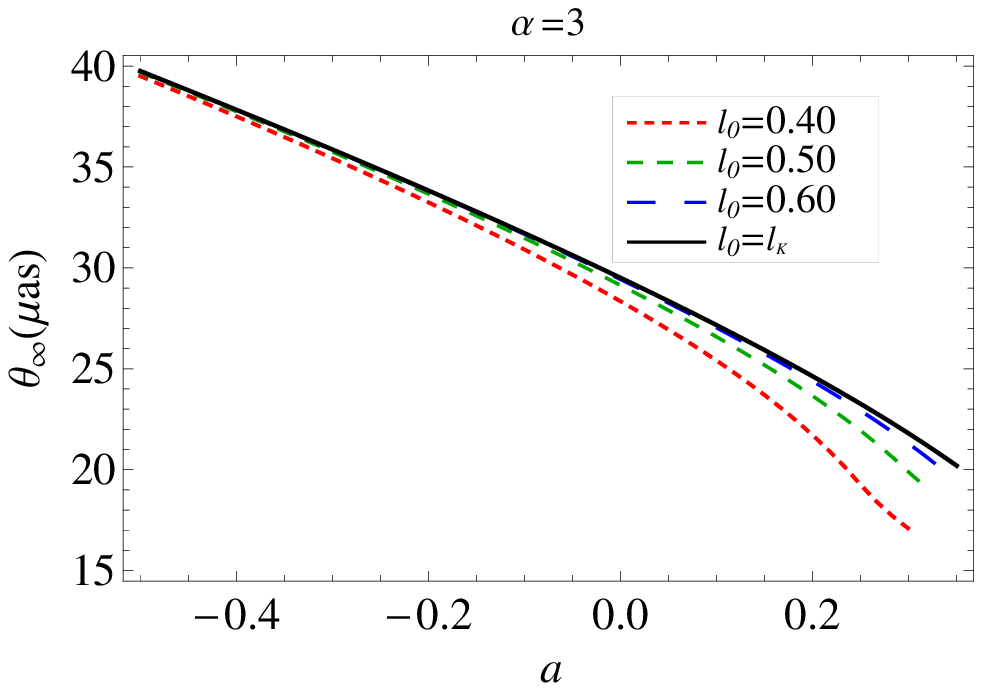}\\
		    \includegraphics[scale=0.8]{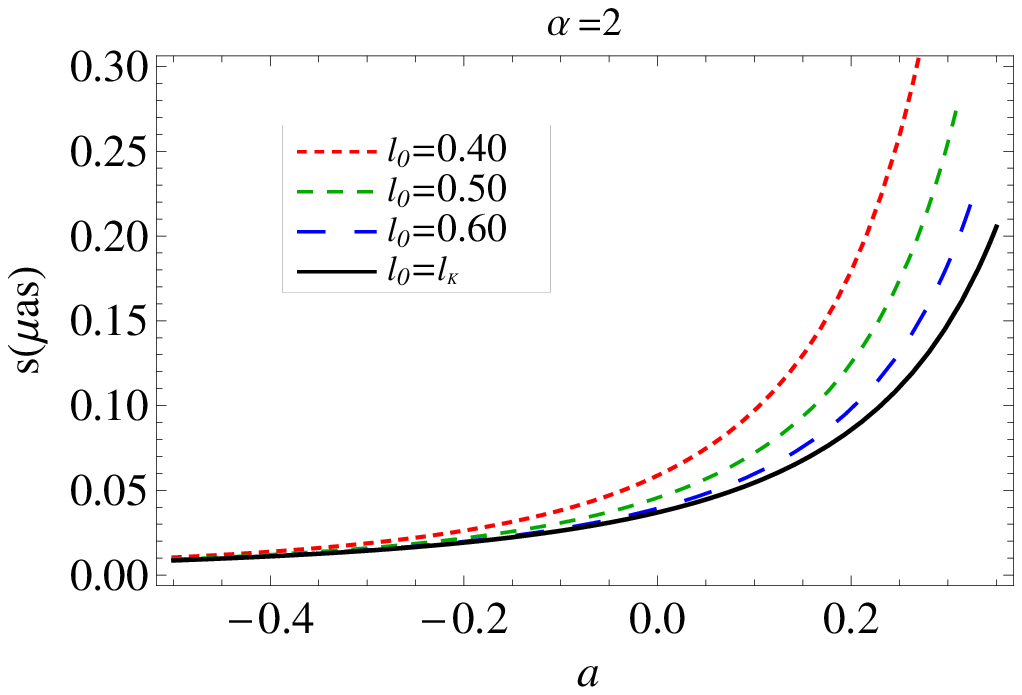}&
		    \includegraphics[scale=0.8]{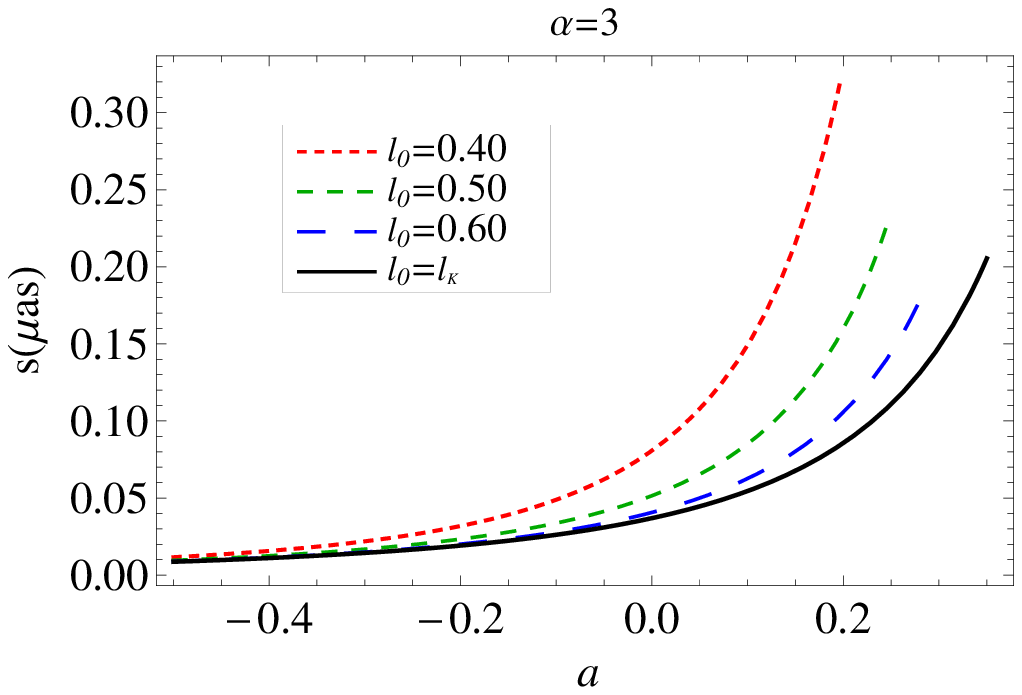}\\
		     \includegraphics[scale=0.8]{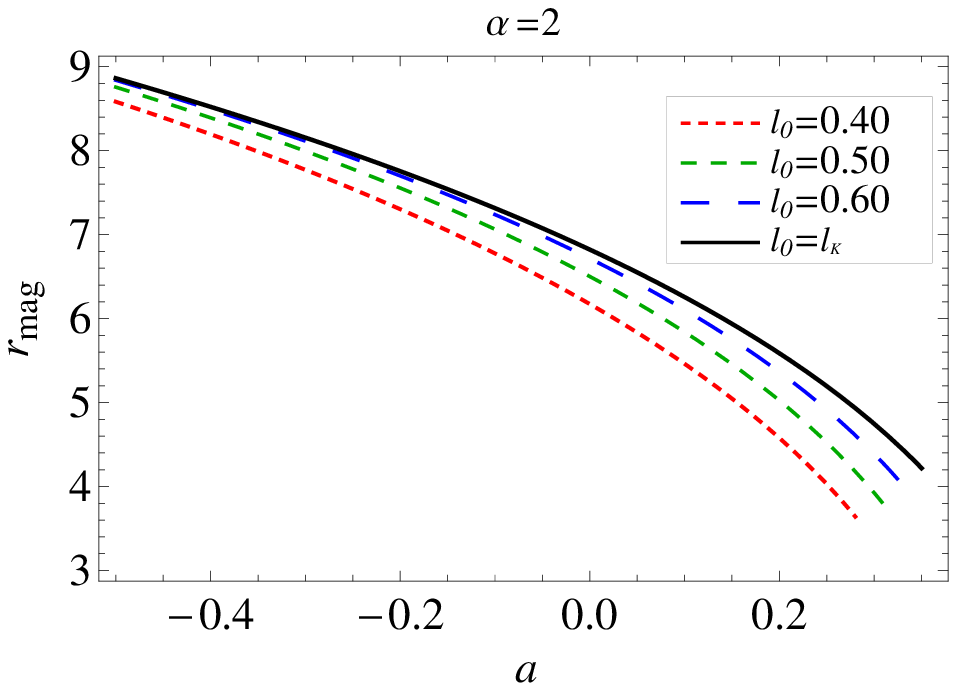}&
		    \includegraphics[scale=0.8]{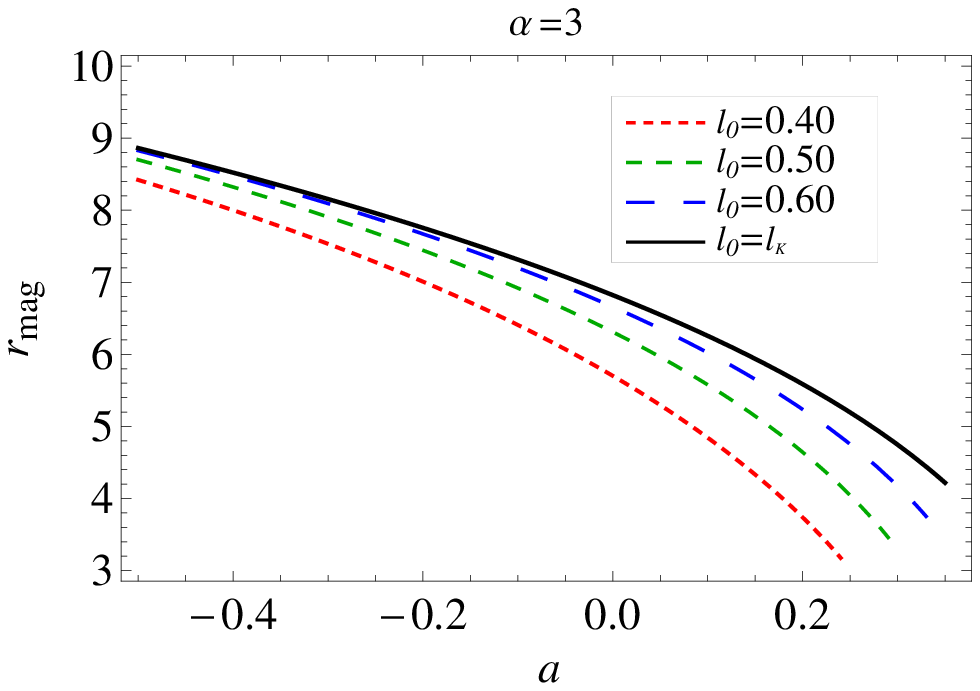}
		 \end{tabular}
	\end{centering}
	\caption{ The behavior of lensing observables [$\theta_{\infty}$ (top), $s$ (middle), and $r_{mag}$ (bottom)] in strong gravitational lensing by hairy Kerr black holes, and their comparison with the Kerr black holes ($\ell_{0}=\ell_{K}$), by taking the  Sgr A* black hole as lens. If Sgr A* is considered to be the  Kerr black hole ($a=0.1$), then $\theta_{\infty,\text{Kerr}}=24.23~\mu$as, $s_{\text{Kerr}}=48.64~$nas, $r_{mag,\text{Kerr}}=6.2$, and $\Delta T_{2,1,\text{Kerr}}=10.58~min$.}\label{plot7}		
\end{figure}  

Besides the position, magnification of the images can be  another good source of information. The brightness of the relativistic images  will be magnified by the lensing. Classically, magnification is the  ratio of the  angular area element of the image and the corresponding area element of the unlensed source. For $n$-loop relativistic images magnification  is given by \cite{Bozza:2002zj,Bozza:2002af}
\begin{eqnarray}
\mu_n &=& \frac{1}{\beta} \Bigg[\frac{u_m}{D_{OL}}(1+e_n) \Bigg(\frac{D_{OL}+D_{LS}}{D_{LS}}\frac{u_me_n}{D_{OL} \bar{a}}  \Bigg)\Bigg].
\end{eqnarray}
Thus, magnification decreases exponentially  with $n$ and the images become fainter.

\begin{table}[!htbp]
 \resizebox{\textwidth}{!}{ 
 \begin{centering}
	\begin{tabular}{p{1cm}  p{1.5cm} p{1.6cm} p{1.6cm} p{2.2cm} p{1.5cm} p{1.6cm} p{1.6cm} p{2.2cm} p{1cm} }
\hline\hline
\multicolumn{1}{c}{}&
\multicolumn{4}{c}{Sgr A*}&
\multicolumn{4}{c}{M87*}& \\
{$a$ }  & {$\theta_\infty $ ($\mu$as)} & {$s$ (nas) } & $\Delta T_{2,1}$(min)& $\Delta \widetilde{T}_{1,1}$(min) & {$\theta_\infty $ ($\mu$as)} & {$s$ (nas) } &$\Delta T_{2,1}$(hrs)& $\Delta \widetilde{T}_{1,1}$(hrs) & {$r_m $ } 
\\ \hline
\hline
\\

-0.2&30.1881 & 17.088 & 13.1814 & 5.36668 & 22.6808 & 12.8385 & 332.091 & 135.207 & 7.75544 \\
\hline \\ 
-0.1&28.3036 & 23.3341 & 12.3586 & 2.58598 & 21.2649 & 17.5312 & 311.359 & 65.1506 & 7.3151 \\
\hline \\ 
0 &26.3299 & 32.9517 & 11.4968 & 0. & 19.782 & 24.7571 & 0. & 289.647 & 6.82188 \\
\hline \\ 
0.1&24.2376 & 48.6456 & 10.5832 & -2.58598 & 18.2101 & 36.5481 & 266.631 & -65.1506 & 6.25687 \\
\hline \\ 
0.2&21.977 & 76.4153 & 9.5961 & -5.36668 & 16.5116 & 57.412 & 241.762 & -135.207 & 5.58737 \\

\hline
\end{tabular}
\end{centering}}
\caption{The lensing observables of two consecutive images on the same side for the  Kerr black hole ($\alpha=0$) compared with the Schwarzschild black hole ($a=0$), considering supermassive black holes Sgr A* and M87* as the lens. The $\Delta \widetilde{T}_{1,1}$  corresponds to the time delay between prograde and retrograde images of the same order.
\label{table1} }
\end{table}  

\begin{table}[!htbp]
 \resizebox{\textwidth}{!}{ 
 \begin{centering}
	\begin{tabular}{p{0.8cm} p{0.8cm} p{1.5cm} p{1.5cm} p{2cm} p{2cm} p{1.5cm} p{1.5cm} p{2cm} p{2cm} p{1cm} }
\hline\hline
\multicolumn{2}{c}{}&
\multicolumn{4}{c}{Sgr A*}&
\multicolumn{4}{c}{M87*}& \\
{$a$ } & {$\ell_0$} & {$\theta_\infty $ ($\mu$as)} & {$s$ (nas) } &$\Delta T_{2,1}$(min) & $\Delta \widetilde{T}_{1,1}$(min) & {$\theta_\infty $ ($\mu$as)} & {$s$ (nas) } &$\Delta T_{2,1}$(hrs)& $\Delta \widetilde{T}_{1,1}$(hrs) & {$r_m $ } 
\\ \hline
\hline
\\

\multirow{3}{*}{-0.2} 

& 0.40 & 29.859 & 23.389 & 13.0377 & 6.71112 & 22.4335 & 17.5725 & 328.47 & 169.079 & 7.30294 \\
& 0.50 & 30.0891 & 19.428 & 13.1382 & 5.95866 & 22.6064 & 20.6399 & 331.002 & 150.121 & 7.55538 \\
& 0.60 & 30.1708 & 17.6458 & 13.1739 & 5.55576 & 22.6678 & 18.3293 & 331.901 & 139.971 & 7.69973 \\
\hline \\

\multirow{3}{*}{-0.1} 

&0.40& 27.8565 & 34.0835 & 12.1634 & 3.08272 & 20.929 & 25.6074 & 306.442 & 77.6653 & 6.77644 \\
&0.50& 28.1611 & 27.4717 & 12.2964 & 2.8013 & 21.1578 & 20.6399 & 309.792 & 70.5755 & 7.06377 \\
&0.60& 28.2762 & 24.3963 & 12.3466 & 2.65115 & 21.2443 & 18.3293 & 311.058 & 66.7924 & 7.23876 \\

\hline \\

\multirow{3}{*}{0.0}

& 0.40 & 25.7069 & 52.3657 & 11.2248 & 0. & 19.314 & 39.3431 & 282.794 & 0. & 6.17328 \\
& 0.50 & 26.1191 & 40.6789 & 11.4047 & 0. & 19.6237 & 30.5626 & 287.329 & 0. & 6.50177 \\
& 0.60 & 26.2852 & 35.0851 & 11.4772 & 0. & 19.7484 & 26.36 & 289.155 & 0. & 6.71566 \\

\hline \\

\multirow{3}{*}{0.1}   
&0.40 &23.3373 & 86.5985 & 10.1901 & -3.08272 & 17.5337 & 65.0627 & 256.727 & -77.6653 & 5.46158 \\
&0.50 &23.9138 & 64.2125 & 10.4418 & -2.8013 & 17.9668 & 48.2438 & 263.069 & -70.5755 & 5.84032 \\
&0.60 &24.1614 & 53.2539 & 10.5499 & -2.65115 & 18.1528 & 40.0105 & 265.792 & -66.7924 & 6.1056 \\
\hline\\

\multirow{3}{*}{0.2} 

& 0.40 & 20.5986 & 159.964 & 8.99426 & -6.71112 & 15.4761 & 120.184 & 226.6 & -169.079 & 4.57827 \\
& 0.50 & 21.4494 & 111.533 & 9.36575 & -5.95866 & 16.1153 & 48.2438 & 235.959 & -150.121 & 5.02474 \\
& 0.60 & 21.8385 & 87.4929 & 9.53566 & -5.55576 & 16.4076 & 40.0105 & 240.239 & -139.971 & 5.36345 \\

\hline\hline
	\end{tabular}
\end{centering}}
\caption{The lensing observables of two consecutive images on the same side for hairy black holes ($\alpha=2$)  compared with the hairy Schwarzschild black hole ($a=0$), considering supermassive black holes Sgr A* and M87* as the lens. The $\Delta \widetilde{T}_{1,1}$  corresponds to the time delay between prograde and retrograde images of the same order.
\label{table2} }
\end{table}

\begin{table}[!htbp]
 \resizebox{\textwidth}{!}{ 
 \begin{centering}
	\begin{tabular}{p{0.8cm} p{0.8cm} p{1.5cm} p{1.5cm} p{2cm} p{2cm} p{1.5cm} p{1.5cm} p{2cm} p{2cm} p{1cm} }
\hline\hline
\multicolumn{2}{c}{}&
\multicolumn{4}{c}{Sgr A*}&
\multicolumn{4}{c}{M87*}& \\
{$a$ } & {$\ell_0$} & {$\theta_\infty $ ($\mu$as)} & {$s$ (nas) } &$\Delta T_{2,1}$(min) & $\Delta \widetilde{T}_{1,1}$(min) & {$\theta_\infty $ ($\mu$as)} & {$s$ (nas) } &$\Delta T_{2,1}$(hrs)& $\Delta \widetilde{T}_{1,1}$(hrs) & {$r_m $ } 
\\ \hline
\hline
\\

\multirow{3}{*}{-0.2} 
&0.40&29.6646 & 28.4169 & 12.9529 & 8.49563 & 22.2875 & 21.3501 & 326.332 & 214.037 & 7.00919 \\
&0.50&30.0363 & 20.8352 & 13.1152 & 6.42596 & 22.5667 & 22.6087 & 330.421 & 161.894 & 7.44385 \\
&0.60&30.1621 & 17.9379 & 13.1701 & 5.66669 & 22.6612 & 18.7548 & 331.804 & 142.765 & 7.67104 \\

\hline \\

\multirow{3}{*}{-0.1} 
&0.40 & 27.5793 & 43.5975 & 12.0423 & 3.58273 & 20.7208 & 32.7555 & 303.392 & 90.2626 & 6.40824 \\
&0.50 & 28.0833 & 30.0921 & 12.2624 & 2.94887 & 21.0994 & 22.6087 & 308.937 & 74.2932 & 6.91945 \\
&0.60 & 28.2622 & 24.9626 & 12.3405 & 2.68718 & 21.2338 & 18.7548 & 310.904 & 67.7004 & 7.199 \\
\hline \\

\multirow{3}{*}{0.0}
&0.40& 25.2927 & 72.0643 & 11.0439 & 0. & 19.0028 & 54.143 & 278.238 & 0. & 5.70217 \\
&0.50& 26.0003 & 45.932 & 11.3529 & 0. & 19.5344 & 34.5094 & 286.022 & 0. & 6.31063 \\
&0.60& 26.262 & 36.2528 & 11.4671 & 0. & 19.731 & 27.2373 & 288.901 & 0. & 6.6594 \\
\hline \\

\multirow{3}{*}{0.1}   
&0.40 & 22.6713 & 133.063 & 9.89926 & -3.58273 & 17.0333 & 99.9726 & 249.4 & -90.2626 & 4.84366 \\
&0.50 & 23.7215 & 75.9211 & 10.3578 & -2.94887 & 17.8223 & 57.0406 & 260.953 & -74.2932 & 5.57822 \\
&0.60 & 24.1211 & 55.8805 & 10.5323 & -2.68718 & 18.1225 & 41.9839 & 265.349 & -67.7004 & 6.02345 \\
\hline\\

\multirow{3}{*}{0.2} 
&0.40 & 19.3775 & 294.298 & 8.46106 & -8.49563 & 14.5586 & 221.11 & 213.166 & -214.037 & 3.73983 \\
&0.50 & 21.1066 & 142.375 & 9.21605 & -6.42596 & 15.8577 & 57.0406 & 232.187 & -161.894 & 4.64357 \\
&0.60 & 21.7627 & 94.2356 & 9.50255 & -5.66669 & 16.3507 & 41.9839 & 239.405 & -142.765 & 5.237 \\
\hline\hline
	\end{tabular}
\end{centering}}
\caption{The lensing observables of two consecutive images on the same side for hairy black holes ($\alpha=3$)   compared with the hairy Schwarzschild black hole ($a=0$), considering supermassive black holes Sgr A* and M87* as the lens. The $\Delta \widetilde{T}_{1,1}$  corresponds to the time delay between prograde and retrograde images of the same order.
\label{table3} }
\end{table}  

\subsection{Time delay}
The time delay is an important observable in strong field lensing, which is defined as the time lag between the formation of relativistic images. The deflection angle for  hairy black holes could be more than $2\pi$, and multiple images of the source $S$ can be formed. The time travelled by the light paths corresponding to the different  images is not the same and hence there is a time difference between the two images. We can   determine  the time delay between the relativistic images using the recipe of  Bozza and Manchini \cite{Bozza:2003cp}. The prerequisite for measuring the time delay is assuming the source with luminosity variations, which would show up in the images with a temporal phase depending upon the geometry of the lens. Due to the dimensional variability, time delay is useful to determine the length scale and mass of lensing system  and   in cosmological contexts  it is possible to determine the Hubble parameter \cite{Refsdal:1964,Blandford:1992,Walsh:1979}. The time delay between the $p$th and $q$th image, when they are on the same side of the lens, can be approximated as  \cite{Bozza:2003cp}
\begin{eqnarray}\label{td}
\Delta T_{p,q} \approx 2\pi(p-q) \frac{\widetilde{R}(0,x_m)}{\bar{a}\sqrt{{c_2}_m}} + 2\sqrt{\frac{A_m u_m}{B_m}}\left[ e^{(\bar{b}-2q\pi\pm \beta)/2\bar{a}}-e^{(\bar{b}-2p\pi\pm \beta)/2\bar{a}}\right] 
\end{eqnarray}

where 
\begin{eqnarray}
\widetilde{R}(z,x_m) &=& \frac{2 x^2 \sqrt{B(x)A(x_0)}[C(x)-L D(x)]}{x_0 \sqrt{C(x)(D(x)^2+ A(x)C(x))}}\left(1- \frac{1}{\sqrt{A(x_0)}f(z,x_0)}\right)
\end{eqnarray}
The maximum contribution to the time delay in Equation~(\ref{td}) comes from the first term  while the second term has a negligible contribution.  Thus, the time delay between the $p$th and $q$th image becomes \cite{Bozza:2003cp}
\begin{eqnarray}\label{td2}
\Delta T_{p,q} \approx 2\pi(p-q)\frac{\widetilde{R}(0,x_m)}{\bar{a}\sqrt{{c_2}_m}}=2\pi(p-q)u_m
\end{eqnarray}
The time delay for direct photons ($a>0$) is different than the prograde photons ($a<0$)\cite{Bozza:2003cp}. To distinguish from the case  above when the two images are on the same side of the lens, we now have the case when the two images are opposite sides of the lens. Then the time delay reads \cite{Bozza:2003cp}
\begin{eqnarray}\label{tdo}
\Delta \widetilde{T}_{p,q} \approx \frac{\tilde{a}(a)}{\bar{a}(a)}[2\pi p -\bar{b}(a)]+\tilde{b}(a)-\frac{\tilde{a}(-a)}{\bar{a}(-a)}[2\pi q -\bar{b}(-a)]+\tilde{b}(-a),
\end{eqnarray}
where
\begin{eqnarray}
\tilde{a} = \frac{\widetilde{R}(0,x_m)}{\sqrt{{c_2}_m}}~~ \text{and}~~ \tilde{b} = -\pi + \tilde{I}_R(x_m) + \tilde{a}~\text{log}\left( \frac{2 {c_2}_m C_m}{u_m A_m (D_m-J_m A_m)}\right),\\
\tilde{I}_R(x_m) = \int_{0}^{1} [\widetilde{R}(z,x_m)f(z,x_m)-\widetilde{R}(0,x_m)f_0(z,x_m)]dz.
\end{eqnarray}

The time delay for  Sgr A* and M87* black holes are estimated and tabulated in Tables~\ref{table1}--\ref{table3}. As can be easily verified from these tables, the time delay for the images when they are on the same side as that of lens $\Delta T_{2,1}$, as expected,  decreases with spin and $\alpha$ but increases with $\ell_0$. However, the situation is different when images are on the opposite side, as the time delay $\Delta \widetilde{T}_{1,1}$ is negative for  $a>0$, suggesting that the direct photons take lesser time to encircle the black hole than the retrograde photons. 

If the first image can be distinguished from the other inner packed ones, we can have three distinguishable observables \cite{Bozza:2002zj}, given by
\begin{eqnarray}
\theta_\infty &=& \frac{u_m}{D_{OL}},\\
s &=& \theta_1-\theta_\infty \approx \theta_\infty ~\text{exp}\left({\frac{\bar{b}}{\bar{a}}-\frac{2\pi}{\bar{a}}}\right),\\
r_{\text{mag}} &=& \frac{\mu_1}{\sum{_{n=2}^\infty}\mu_n } \approx 
\frac{5 \pi}{\bar{a}~\text{log}(10)},
\end{eqnarray} 
where $\theta_\infty$ is angular position of the images obtained and  $s$ and $r_{\text{mag}}$  are, respectively, the angular separation and the difference in magnitude of the flux between the first image  and other packed images. It is worth it to note that, for hairy Kerr black holes, the angular position and the relative magnification decrease with $\alpha$ and increase with $\ell_0$ but are smaller than the Kerr black hole.  On the other hand, the angular separation decreases with $\ell_0$  but increases with $\alpha$. For retrograde orbits, $s$ is smaller than the prograde orbits. 
\begin{figure}[t]
	\begin{centering}
		\begin{tabular}{ccc}
		    \includegraphics[scale=0.8]{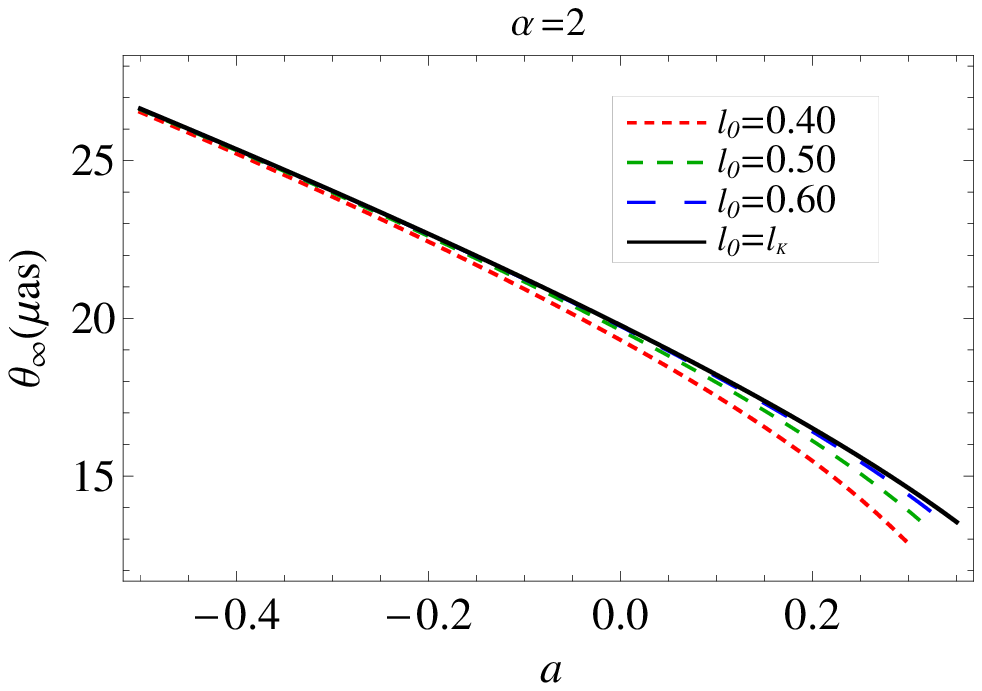}&
		    \includegraphics[scale=0.8]{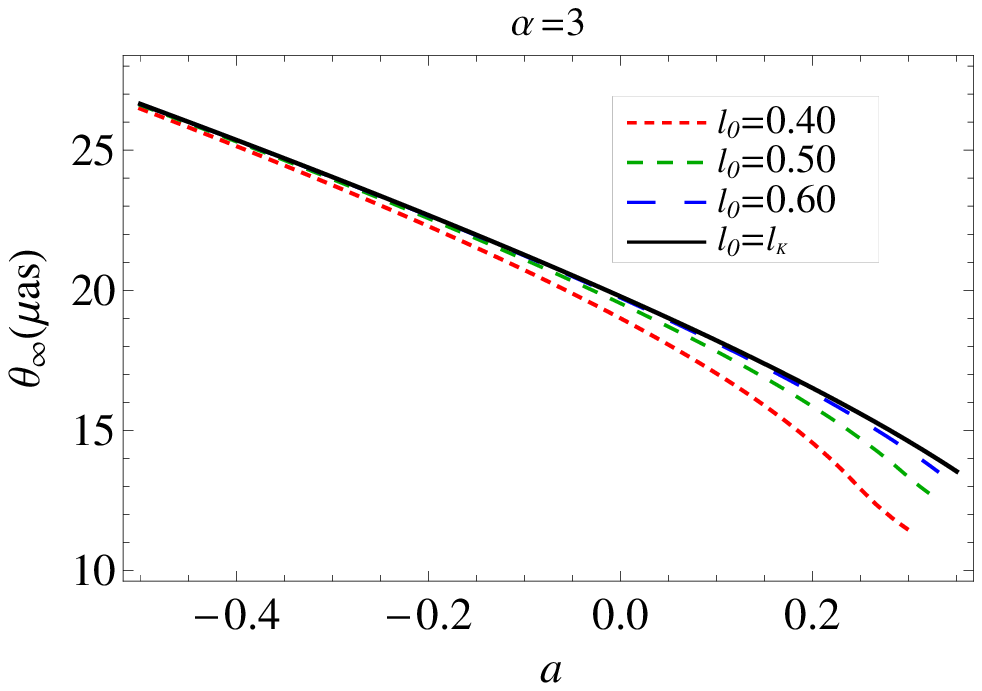}\\
		    \includegraphics[scale=0.8]{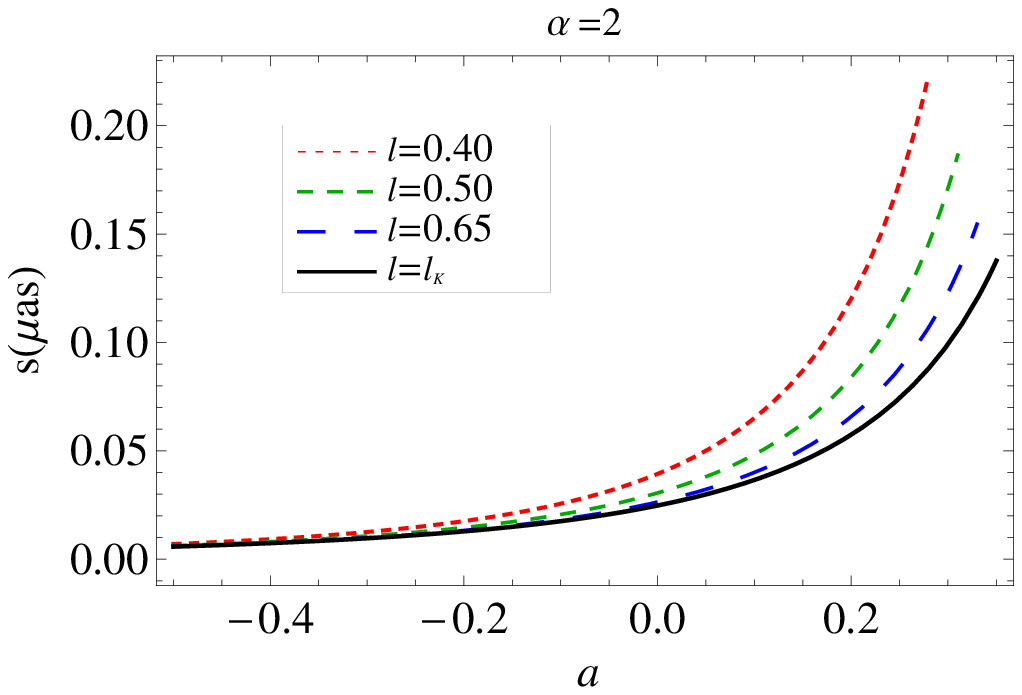}&
		    \includegraphics[scale=0.8]{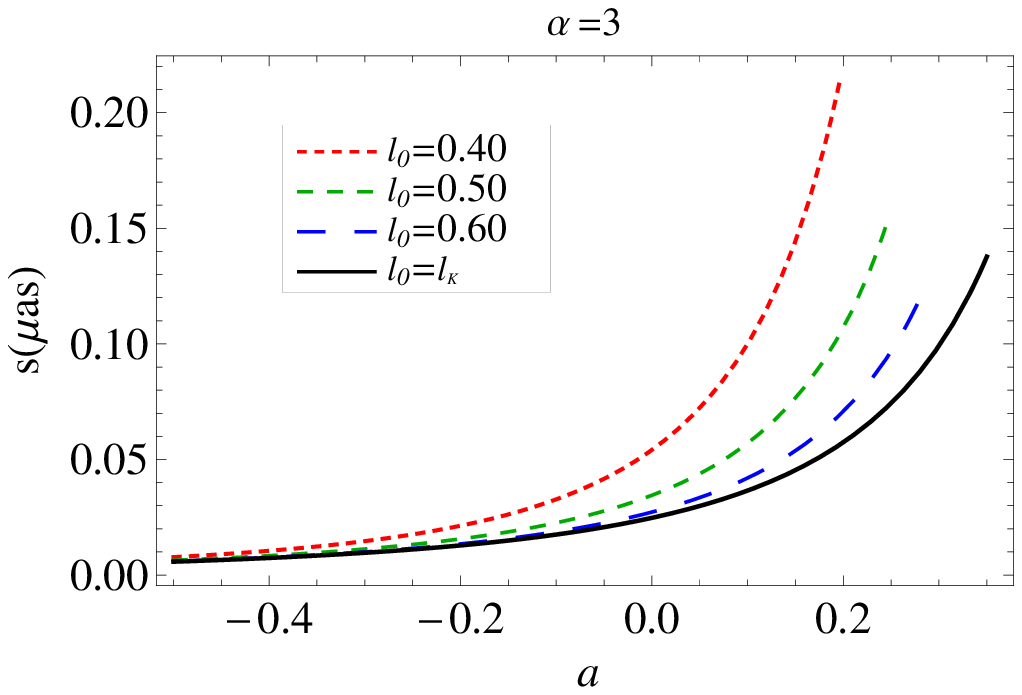}\\
		     \includegraphics[scale=0.8]{mag1.eps}&
		    \includegraphics[scale=0.8]{mag2.eps}
		 \end{tabular}
	\end{centering}
	\caption{ The behavior of the lensing observables [$\theta_{\infty}$ (top), $s$ (middle) and $r_{mag}$ (bottom)] in strong gravitational lensing by hairy Kerr black holes and their comparison with the Kerr black holes ($\ell_{0}=\ell_{K}$) by taking  the M87* black hole as the lens. If  M87* is considered as the Kerr black hole ($a=0.1$), then $\theta_{\infty,\text{Kerr}}=18.21~\mu$as, $s_{\text{Kerr}}=36.54~$nas, $r_{mag,\text{Kerr}}=6.2$, and $\Delta \text{T}_{1,2,\text{Kerr}}=266.63~hrs$.  }\label{plot8}	\end{figure}

\section{Gravitational lensing by Supermassive black holes}\label{cal}
We investigate strong gravitational lensing effects by supermassive black holes Sgr A* and M87*. In turn, we shall compare the lensing results of the Kerr black hole with those of hairy Kerr black holes.  Taking the supermassive black holes Sgr A* and M87* as the lens, respectively, with mass  $M=4.3\times 10^6 M_{\odot}$  and $M=6.5 \times 10^9 M_{\odot} $, and $D_{OL}$, respectively, as 8.35~Kpc~\cite{Do:2019vob} and 16.8~Mpc~\cite{Akiyama:2019cqa}, we numerically calculate the observables $\theta_{\infty}$, separation $s$, and relative magnification $r_{\text{mag}}$ of  the hairy Kerr black holes for different values of $\alpha$ and $\ell_0$, and  depict this in Fig.~\ref{plot7} (Sgr A*) and Fig.~\ref{plot8} (M87*). Tables~\ref{table1}--\ref{table3} shows the lensing  observables along with the time delay for various values of $a$, $\alpha$, and $\ell_0$  in comparison  with the Schwarzschild~($a=\alpha=0$) and the Kerr black hole~($\alpha=0$).  We also show the time delay $\Delta \widetilde{T}_{1,1}$ between prograde and retrograde images of the same order in these tables, but other corresponding observables are not tabled. Considering hairy black holes as the lens, we observe that the angular position of images for Sgr A* and M87*  $\theta_{\infty}$, is in the range of $19.37~\mu\text{as} < \theta_{\infty}<26.26~\mu\text{as}$ and  $14.55~\mu\text{as} < \theta_{\infty}< 19.734~\mu\text{as}$, respectively, and the latter are consistent with the EHT measured diameters of M87* shadow $42\pm 3~\mu\text{as}$. However their deviation from the Kerr black hole (as shown in Table~\ref{table4})  are not more than  $2.59~\mu$as and $1.953~\mu$as, respectively,  for $\alpha = 3$. Although they deviate from the Kerr black hole, it is impossible to distinguish the hairy Kerr black holes from the Kerr black hole using the currently available observation facility of the EHT. Further, the angular separation $s$ between the first and other packed images due to the hairy Kerr black hole for Sgr A* and M87* are in the range of $36~\text{nas} <s<294~$nas and   $27~\text{nas} <s<222~$nas, , respectively (cf. Table~\ref{table3}), while their deviations are $|\delta s|\le 217.8~$nas and $|\delta s|\le 163.7~$nas, respectively, which are beyond the threshold of the current  EHT observation. We may have to wait for the next generation event horizon telescope (ngEHT) for this purposes. For given values of parameters $a$, $\alpha$, and $\ell_0$, the angular position and angular separation of the relativistic images for the Sgr A* black hole are larger than the M87* black hole. The time delay of the first image from that of the second image, $\Delta T_{2,1}$, for the hairy Kerr black holes  Sgr A* and M87* can reach as much as $11.46$~min and $288.9$~hrs, respectively, (cf. Table~\ref{table3}), while the deviation from the Kerr black hole for Sgr A* and M87* is $1.13$~min and $28.6$~hrs, respectively. On the other hand the time delay $\widetilde{T}_{1,1}$ can reach $8.49~$min for Sgr A* and $214.03$~hrs for M87*, while the deviation is $3.12$~min and $78.83$~hrs, respectively. Thus, the time delay in Sgr A* is much shorter for observation and more difficult for measurement. In the case of M87* the time delay  $\Delta T_{2,1}$ can reach to the order of a few hundred hours, and their deviation from those of Kerr black holes can reach up to tens of hours. These are sufficient times for astronomical measurements, provided we have enough angular resolution separating two  relativistic images.  Presently, this does  not seem feasible, while $\Delta \widetilde{T}_{1,1}$ could be easier to measure since the two images would be on opposite sides of the black hole and could be resolved by current interferometry.   

\begin{table}[!htbp]
 \resizebox{\textwidth}{!}{ 
 \begin{centering}
	\begin{tabular}{p{0.8cm} p{0.8cm} p{1.5cm} p{1.5cm} p{2cm} p{2cm} p{1.5cm} p{1.5cm} p{2cm} p{2cm} p{1cm} }
\hline\hline
\multicolumn{2}{c}{}&
\multicolumn{4}{c}{Sgr A*}&
\multicolumn{4}{c}{M87*}& \\
{$a$ } & {$\ell_0$} & {$\delta\theta_\infty $ ($\mu$as)} & {$\delta s$ (nas) } &$\delta \Delta T_{2,1}$(min)& $\delta \Delta \widetilde{T}_{1,1}$(min) & {$\delta\theta_\infty $ ($\mu$as)} & {$\delta s$ (nas) } & $\delta\Delta T_{2,1}$(hrs) &$\delta\Delta \widetilde{T}_{1,1}$(hrs) & {$\delta r_m $ } 
\\ \hline
\hline
\\

\multirow{3}{*}{-0.2} 
&0.40&0.523501 & -11.3289 & 0.228584 & -3.12895 & 0.393314 & -8.51161 & 5.75889 & -78.8302 & 0.746251 \\
&0.50& 0.151781 & -3.74725 & 0.0662744 & -1.05928 & 0.114036 & -2.81536 & 1.6697 & -26.6873 & 0.311598 \\
&0.60& 0.0260408 & -0.849909 & 0.0113705 & -0.30001 & 0.0195648 & -0.63855 & 0.286467 & -7.5584 & 0.0844038 \\
\hline \\

\multirow{3}{*}{-0.1} 
&0.40& 0.724239 & -20.2635 & 0.316234 & -0.996753 & 0.544132 & -15.2242 & 7.96715 & -25.112 & 0.906862 \\
&0.50& 0.22022 & -6.75809 & 0.0961575 & -0.362891 & 0.165454 & -5.07745 & 2.42257 & -9.14259 & 0.395656 \\
&0.60& 0.0413634 & -1.6285 & 0.0180611 & -0.101208 & 0.031077 & -1.22352 & 0.455027 & -2.5498 & 0.116096 \\
\hline \\

\multirow{3}{*}{0.0}
&0.40& 1.03716 & -39.1126 & 0.45287 & 0. & 0.779235 & -29.3859 & 11.4095 & 0. & 1.11972 \\
&0.50& 0.329592 & -12.9803 & 0.143914 & 0. & 0.247628 & -9.75229 & 3.62575 & 0. & 0.511247 \\
&0.60& 0.0678679 & -3.30112 & 0.0296341 & 0. & 0.0509902 & -2.48018 & 0.746596 & 0. & 0.162478 \\
\hline \\

\multirow{3}{*}{0.1}   
&0.40& 1.56634 & -84.4179 & 0.683932 & 0.996753 & 1.17681 & -63.4244 & 17.2308 & 25.112 & 1.41322 \\
&0.50& 0.516108 & -27.2755 & 0.225355 & 0.362891 & 0.387759 & -20.4925 & 5.67755 & 9.14259 & 0.678647 \\
&0.60& 0.116494 & -7.23496 & 0.0508662 & 0.101208 & 0.0875234 & -5.43574 & 1.28151 & 2.5498 & 0.23342 \\
\hline\\

\multirow{3}{*}{0.2} 
&0.40& 2.59946 & -217.882 & 1.13504 & 3.12895 & 1.95302 & -163.698 & 28.5959 & 78.8302 & 1.84754 \\
&0.50& 0.870398 & -65.9595 & 0.380054 & 1.05928 & 0.653943 & -49.5563 & 9.575 & 26.6873 & 0.943799 \\
&0.60& 0.214253 & -17.8204 & 0.0935522 & 0.30001 & 0.160972 & -13.3887 & 2.35694 & 7.5584 & 0.350367 \\
\hline\hline
	\end{tabular}
\end{centering}}
\caption{Deviation of the lensing observables of two consecutive images on the same side of hairy Kerr black holes from the Kerr black hole by taking Sgr A* and M87* as the lens  ($\alpha=3$), where $\delta X = X_{\text{Kerr}}-X_{\text{hairy Kerr}}$.  The $\delta\Delta \widetilde{T}_{1,1}$  corresponds to the time delay between prograde and retrograde images of the same order.
\label{table4} }
\end{table}

\section{Conclusions}\label{conclusion}
The EHT Collaboration has recently captured the image of the  supermassive black hole M87* at a 1.3 $mm$ wavelength with an angular resolution of 20 $\mu\text{as}$ \cite{Akiyama:2019cqa}.  Though the observed shadow is consistent with the Kerr black hole's image as predicted by  GR, the observation did not tell about most modified gravity theories or alternatives to the Kerr black hole. Hence they could not be wholly ruled out, and  we can't ignore deviations from the Kerr black hole (e.g. hairy Kerr black holes) due to additional sources or arising as a solution from modified theories of gravity. These hairy black holes, in Boyer-Lindquist coordinates, are defined by the metric~(\ref{metric}) with mass function $\tilde{m}(r)$, and Kerr black holes are included as a particular case when $\tilde{m}(r) = M$ ($\alpha=0$). The impact of the deviation parameter $\alpha$, arising due to surrounding matter on gravitational lensing, presents a good theoretical opportunity to distinguish the hairy rotating black holes from the Kerr black hole and test whether astrophysical black hole candidates are the black holes as predicted by Einstein's GR. Motivated by the above arguments, we have examined the effects of $\alpha$ and $\ell_0$, in a strong field observation on the lensing observables due to rotating hairy black holes and compared with those due to the Kerr black holes. We have numerically calculated the strong lensing coefficients and lensing observables as functions of $\alpha$ for relativistic images. In turn, we have applied our results to the supermassive black holes Sgr A* and M87* at the centre of galaxies.  Our analysis shows that such hairy Kerr black hole properties are qualitatively different from the Kerr black hole.

We highlight results that are obtained by our analysis.  The horizon radius of the hairy black hole increases with $l_0$ and coincides with the maximum horizon radius of the Kerr black hole in the limit $l_0 \to 1$, whereas the horizon radius  decreases with $\alpha$~(cf. Fig~\ref{plot1}). This is also true for the photon sphere. Interestingly, the lensing coefficient $\bar{a}$, like the Kerr black hole case, increases with $a$ whereas $\bar{b}$ decreases. $\bar{a}$, for hairy Kerr black holes, always takes a larger value when compared with the Kerr black hole  and $\bar{b}$ is smaller (cf. Fig.~\ref{plot4}). Both $\bar{a}$ and $\bar{b}$ diverge with opposite sign at critical values of $a$ (e.g., for $\ell_0=0.4$ and  $\alpha=2 , a=0.39$). This signals that the strong deflection angle is no longer valid. The deflection angle $\alpha_D$, increases with $u$ and diverges at impact parameter $u=u_m$. $u_m$ decreases with with $\ell_0$ as well as $\alpha$ ~(cf. Fig.~\ref{plot5}). 

We have also  numerically  calculated lensing observables $\theta_{\infty}$, separation $s$,  and relative magnitude $r_{mag}$.  $\theta_{\infty}$ and $r_{mag}$ decrease with the increasing $a$. They take smaller values compared to the Kerr black hole and both decrease  with the deviation parameter $\alpha$ (cf. Figs.~\ref{plot7} and ~\ref{plot8} and Tables~\ref{table1}--\ref{table3}). Similarly, the observable $s$ increases with $a$ as well as with $\alpha$, which means that separation between two relativistic images for hairy Kerr black holes is larger than the analogous Kerr black hole. Finally, considering the supermassive black holes Sgr A* and M87* as hairy Kerr black holes, we have analyzed the magnitude of lensing observables. It turns out that, for given values of parameters $a$, $\alpha$ and  $\ell_0$, the images of Sgr A* are less packed than M87*.

 After calculating the observables for Sgr A* and M87* in the strong deflection limit, we found that the observables of hairy Kerr black holes are different from that of Kerr black holes. However, based on strong gravitational lensing observables and relativistic images, it is difficult to distinguish the hairy black holes from the Kerr black hole, at least from the presently available astronomical observations like the EHT, and we may have to wait for ngEHT for appropriate resolutions.   Therefore our results, in principle, could provide a possibility to test how hairy black holes deviate from the Kerr black hole in future astronomical observations. The results presented here are the generalization of previous discussions on the Kerr black holes to a more general setting. The possibility of a different conception of these result to the weak-field gravitational lensing is an interesting problem for future research.

\section{Acknowledgments} S.G.G. and  S.U.I.  would like to thank SERB-DST for the ASEAN project IMRC/AISTDF/CRD/2018/000042.  S.G.G. would also like to thank Rahul Kumar for fruitful discussion.


\begin{thebibliography}{}

\bibitem{Kerr:1963ud} 
R.~P.~Kerr,
Phys.\ Rev.\ Lett.\  {\bf 11}, 237  (1963).

\bibitem{Israel:1967}
W. Israel, Phys. Rev. {\bf 164}, 1776 (1967); 
W. Israel, Commun. Math. Phys. {\bf 8}, 245 (1968).
 
\bibitem{Carter:1971}
B. Carter, Phys. Rev. Lett. {\bf 164}, 331 (1971); 
B. Carter, \textit{Black Hole Equilibrium States}, 
edited by B.S. DeWitt and C. DeWitt (Gordon and Breach, New York, 1973),  p. 57-210.

\bibitem{Robin:1975}
D. C. Robinson, Phys. Rev. Lett. {\bf 34}, 905 (1975).
 
\bibitem{Kumar:2020yem}
R.~Kumar, A.~Kumar and S.~G.~Ghosh,
Astrophys. J. \textbf{896}, 89 (2020). 
 
\bibitem{Ryan:1995}
F.~D.~Ryan,
Phys. Rev. D \textbf{52}, 5707 (1995).


\bibitem{Will:2006}
C.~M.~Will,
Living Rev. Relativity. \textbf{9}, 3 (2006).

\bibitem{Bambi:2013ufa}
C.~Bambi and L.~Modesto,
Phys. Lett. B \textbf{721}, 329 (2013).

\bibitem{Ghosh:2014pba} 
S.~G.~Ghosh,
Eur. Phys. J. C {\bf 75}, 532 (2015).


\bibitem{Azreg-Ainou1:2014pra}
M.~Azreg-Ainou,
Phys. Rev. D {\bf 90}, 064041 (2014).

\bibitem{Ovalle:2017fgl}
J.~Ovalle,
Phys. Rev. D \textbf{95}, 104019 (2017).

\bibitem{Ovalle:2019qyi}
J.~Ovalle,
Phys. Lett. B \textbf{788}, 213 (2019);
J. Ovalle, Mod. Phys. Lett. A \textbf{23}, 3247 (2008).

\bibitem{Contreras:2021yxe}
E.~Contreras, J.~Ovalle, and R.~Casadio,
Phys. Rev. D \textbf{103}, 044020 (2021).

\bibitem{Herdeiro:2015waa}
C.~A.~R.~Herdeiro and E.~Radu,
Int. J. Mod. Phys. D \textbf{24}, 1542014 (2015).

\bibitem{Herdeiro:2014goa}
C.~A.~R.~Herdeiro and E.~Radu,
Phys. Rev. Lett. \textbf{112}, 221101 (2014).

\bibitem{yaun:2021}
Y. Xing Gao and Y. Xie
Phys. Rev. D \textbf{103}, 043008 (2021).

\bibitem{Herdeiro:2016tmi}
C.~Herdeiro, E.~Radu, and H.~R\'unarsson,
Classical  Quantum Gravity \textbf{33}, 154001 (2016).


\bibitem{Bekenstein:1994}
J. D. Bekenstein and R. H. Sanders, 
 Astrophys. J. \textbf{429}, 480 (1994).

\bibitem{Eiroa:2006}
 E. F. Eiroa, 
Phys. Rev. D \textbf{73}, 043002 (2006).

\bibitem{Sarkar:2006}
 K.~Sarkar and A.~Bhadra,
Classical  Quantum Gravity \textbf{23}, 6101 (2006).

\bibitem{Chen:2009} 
S.~B.~Chen and J.~L.~Jing,
Phys. Rev. D \textbf{80}, 024036 (2009).

\bibitem{Kumar:2020sag}
R.~Kumar, S.~U.~Islam, and S.~G.~Ghosh,
Eur. Phys. J. C \textbf{80}, 1128 (2020).

\bibitem{Islam:2020xmy}
S.~U.~Islam, R.~Kumar, and S.~G.~Ghosh,
J.Cosmol. Astropart. Phys. \textbf{09}, 030 (2020).


\bibitem{Virbhadra:2002} 
K. S. Virbhadra and G. F. R. Ellis, 
Phys. Rev.D \textbf{65},  103004 (2002).

\bibitem{Virbhadra:2008} 
K. S. Virbhadra and C. R. Keeton, 
Phys. Rev. D \textbf{77}, 124014 (2008).

\bibitem{Rauch :1994qd}
K. P. Rauch  and R. D. Blandford,
 Astrophys. J. \textbf{46}, 421 (1994).

\bibitem{Vazquez:2003zm}
S.~E.~Vazquez and E.~P.~Esteban,
Nuovo Cimento B \textbf{119}, 489 (2004).
 
\bibitem{Bozza:2008mi}
V.~Bozza,
Phys. Rev. D \textbf{78}, 063014 (2008).

\bibitem{Bozza:2009yw}
V.~Bozza,
Gen. Relativ. Gravit. \textbf{42}, 2269 (2010).


\bibitem{Ghosh:2020spb}
S.~G.~Ghosh, R.~Kumar, and S.~U.~Islam,
J.Cosmol. Astropart. Phys. \textbf{03}, 030 (2021).

\bibitem{Wei:2011nj}
S.~W.~Wei, Y.~X.~Liu, C.~E.~Fu, and K.~Yang,
J.Cosmol. Astropart. Phys. {\bf 10}, 053 (2012).

\bibitem{Beckwith:2004ae}
K.~Beckwith and C.~Done,
Mon. Not. R. Astron. Soc. \textbf{359}, 1217 (2005).


\bibitem{Hsiao:2019ohy}
Y.~W.~Hsiao, D.~S.~Lee, and C.~Y.~Lin,
Phys. Rev. D \textbf{101},  064070 (2020).

\bibitem{Kapec:2019hro}
D.~Kapec and A.~Lupsasca,
Classical  Quantum Gravity \textbf{37}, 015006 (2020).

\bibitem{Gralla:2019drh}
S.~E.~Gralla and A.~Lupsasca,
Phys. Rev. D \textbf{101},  044031 (2020).

\bibitem{James:2015yla}
O.~James, E.~von Tunzelmann, P.~Franklin, and K.~S.~Thorne,
Classical  Quantum Gravity \textbf{32},  065001 (2015).


\bibitem{Cunha:2019ikd}
 V. P.~Cunha, C.~A.~R.~Herdeiro, and E.~Radu,
Universe \textbf{5}, 220 (2019).

\bibitem{Cunha:2019dwb}
P.~V.~P.~Cunha, C.~A.~R.~Herdeiro, and E.~Radu,
Phys. Rev. Lett. \textbf{123}, 011101 (2019).

\bibitem{Cunha:2016bpi}
P.~V.~P.~Cunha, C.~A.~R.~Herdeiro, E.~Radu, and H.~F.~Runarsson,
Int. J. Mod. Phys. D \textbf{25},  1641021 (2016).

\bibitem{Cunha:2015yba}
P.~V.~P.~Cunha, C.~A.~R.~Herdeiro, E.~Radu, and H.~F.~Runarsson,
Phys. Rev. Lett. \textbf{115},  211102 (2015).

\bibitem{Akiyama:2019cqa} 
K.~Akiyama {\it et al.},
Astrophys. J.  {\bf 875}, L1 (2019);
K.~Akiyama {\it et al.},
Astrophys.\ J.\  {\bf 875}, L6 (2019).

\bibitem{Ovalle:2020kpd}
J.~Ovalle, R.~Casadio, E.~Contreras, and A.~Sotomayor,
Phys. Dark Univ. \textbf{31}, 100744 (2021).

\bibitem{Toshmatov:2017zpr}
B.~Toshmatov, Z.~Stuchl\'\i{}k, and B.~Ahmedov,
Phys. Rev. D \textbf{95},  084037 (2017).


\bibitem{Kumar:2020} 
R.~Kumar and S.~G.~Ghosh,
Classical  Quantum Gravity \textbf{38}, 8, (2021);
R.~Kumar, S.~G.~Ghosh, and A.~Wang,
Phys. Rev. D \textbf{100}, 124024 (2019);
R.~Kumar and S.~G.~Ghosh,
Astrophys. J. \textbf{892}, 78 (2020).

\bibitem{Bozza:2002zj} 
V.~Bozza,
Phys. Rev. D {\bf 66}, 103001 (2002).

\bibitem{Chander:1992pc}
S.~Chandershaker,
\textit{The Mathematical Theory of Black Holes}, 
(Oxford University Press. New York, 1992).

\bibitem{Harko:2009xf}
T.~Harko, Z.~Kovacs, and F.~S.~N.~Lobo,
Phys. Rev. D \textbf{79}, 064001 (2009).
 
\bibitem{Eiroa:2004gh}
E.~F.~Eiroa,
Phys. Rev. D \textbf{71}, 083010 (2005).

\bibitem{Darwin:1959}
C. Darwin, Proc. R. Soc. A, \textbf{249}, 180 (1959).
 
 
\bibitem{Tsukamoto:2016jzh}
N.~Tsukamoto,
Phys. Rev. D \textbf{95}, 064035 (2017).

\bibitem{Weinberg:1972} 
S. Weinberg, 
Gravitation and Cosmology: Principles and Applications of the General Theory of Relativity  (Wiley, New York, 1972).

\bibitem{Bozza:2008ev}
V.~Bozza,
Phys. Rev. D \textbf{78}, 103005 (2008).

\bibitem{Oho:1987ev}
H.C. Ohanian, 
Am. J. Phys. \textbf{55}, 428 (1987).


\bibitem{Bozza:2002af}
V.~Bozza,
Phys. Rev. D \textbf{67}, 103006 (2003).

\bibitem{Bozza:2003cp}
V.~Bozza and L.~Mancini,
Gen. Relativ. Gravit. \textbf{36}, 435 (2004).

\bibitem{Refsdal:1964}
S. Refsdal, Mon. Not. R. Astron. Soc. \textbf{128}, 307 (1964).

\bibitem{Blandford:1992}
R.D. Blandford and  R. Narayan,
Annu. Rev. Astron.  Astroph. \textbf{30}, 311 (1992).

\bibitem{Walsh:1979}
D. Walsh, R.F. Carswell, and  R.J. Weymann,
Nature (London) 279, 381 (1979).

\bibitem{Do:2019vob}
T.~Do, G.~Witzel, A.~K.~Gautam, Z.~Chen, A.~M.~Ghez, M.~R.~Morris, E.~E.~Becklin, A.~Ciurlo, M.~Hosek, and G.~D.~Martinez \textit{et al.,}, Astrophys. J. \textbf{882}, L27 (2019).


\end{thebibliography}
\end{document}